# SUMMARY INDICES IN TREATMENT EFFECT ESTIMATION

Danil Fedchenko

Department of Economics, The University of Melbourne

Most recent version: here.

This paper studies the practice of combining multiple outcomes into a summary index to estimate a causal effect. For common estimators and index constructions, the estimate equals a weighted sum of the estimated effects on the components, with weights that are implicit and rarely reported. The paper derives the weights and shows that, for inverse-covariance-weighted indices, they can be negative and unrestricted in magnitude, so the index effect can have the opposite sign to every component effect. The paper proposes two procedures for valid inference on the index effect: a variance estimator that accounts for the data-dependent weights, and a shifted t-test that requires no such correction. Conventional t-tests of the null of no effect remain valid. Contrary to common claims, summary indices do not generally improve power. Three published studies illustrate the results.

Danil Fedchenko: danil.fedchenko@unimelb.edu.au

# 1. INTRODUCTION

How does one measure happiness, mental health, or women's empowerment in order to quantitatively assess the effects of policies on these important yet broad concepts? Empirical researchers often measure such concepts by combining multiple related metrics into a *summary index*. As a result, such summary indices have become common left-hand-side variables in empirical economic analyses, and positive estimated effects on them usually serve as arguments in favor of studied interventions. This article formalizes the construction of summary indices and derives the statistical properties of the resulting treatment-effect estimates, in order to establish when such an interpretation is warranted and to provide guidance both for researchers who construct summary indices and for readers who interpret estimated effects on them.

There are several approaches to constructing summary indices. This paper focuses on a family of procedures characterized by the following steps. First, the components comprising a summary index are oriented in the same direction and converted to a common scale, usually through standardization using the control group. Then, a weighted sum of the standardized components is computed, and the summary index equals that sum, sometimes standardized once more. This general construction scheme gives rise to a host of procedures that differ in whether and how standardization is used, and in what weights are used in the second step. This family can be divided into two subfamilies based on how the weights are computed.

Summary indices constructed by averaging control-group-standardized components are referred to here as scale-normalized (**SN**) summary indices. This approach became popular following Kling, Liebman, and Katz (2007), who averaged control-group-standardized measures of depression, distress, sleep quality, and other metrics into a mental health summary index used to estimate the effect of moving to a safer neighborhood. Some recent studies following this summary index construction include: Chetty et al. (2011), Casey, Glennerster, and Miguel (2012), Finkelstein et al. (2012), Banerjee et al. (2015), Hoynes, Schanzenbach, and Almond (2016), Blattman, Jamison, and Sheridan (2017), Bursztyn, González, and Yanagizawa-Drott (2020), Christensen et al. (2021), Levy (2021), Stantcheva (2021), Braghieri, Levy, and Makarin (2022), Bhatt et al. (2024), and Hawkins et al. (2025).

In another approach, popularized by Anderson (2008) and implemented in the Stata command `swindex` (Schwab et al. (2020)), weights are determined by the inverse of a covariance matrix. These are referred to here as inverse-covariance (**IC**) summary indices. Recent studies following this summary index construction include: Currie et al. (2015), Haushofer and Shapiro (2016), Cantoni et al. (2017), Cantoni et al. (2019), Chen and Yang (2019), Alfonsi et al.

(2020), Allcott et al. (2020a), Asher and Novosad (2020), Baranov et al. (2020a), Allcott, Gentzkow, and Song (2022), Egger et al. (2022), Carranza et al. (2022), Beraja et al. (2023), Ramos-Toro (2023), Kinnan et al. (2024), and Baseler et al. (2025).

Causal effects on these summary indices are estimated using standard econometric techniques: linear regression, instrumental variables, difference-in-differences, and regression discontinuity designs. All of these estimators can be represented as weighted sums of the outcomes with procedure-specific data-dependent weights. Mogstad, Santos, and Torgovitsky (2018) refer to them as IV-like estimators. I adopt the same term here. Therefore, for all IV-like estimators, because both summary index construction and estimation are linear, *the treatment effect on a summary index equals a weighted sum of the treatment effects on its component outcomes*. The intuition is clearest in a two-arm randomized experiment with vector potential outcomes. If the constructed summary index can be written as $s_i = A_n \boldsymbol{y}_i + b_n$ – as is true for the procedures studied here – then the difference in treated and control means of the summary index, $\bar{s}_1 - \bar{s}_0$, equals $A_n(\bar{\boldsymbol{y}}_1 - \bar{\boldsymbol{y}}_0)$; the constant $b_n$ cancels. This paper generalizes that intuition from the difference-in-means estimator to all IV-like estimators.

The weights, $A_n$, however, are implicit in the summary index construction, and reporting them is not common practice; this article derives their explicit expressions. Since the weights define which weighted sum of the component effects is estimated, two summary indices built from the same components but with different weights estimate different parameters, and the choice of construction is a choice of estimand. For **SN** summary indices, the weights are guaranteed to be positive and sum to one; for **IC** summary indices, the weights can be negative, their magnitude is unrestricted, and they do not sum to one. As a result, the interpretation of the estimated effect on an **IC** summary index can be problematic. It need not be positive (negative) even when the treatment positively (negatively) affects every component of the index, and it can exceed in magnitude the effect on every component, because some components receive weights far larger than one in absolute value. Similar problems related to negative weights in decompositions of other popular estimators have recently been pointed out by, e.g., de Chaisemartin and d'Haultfœuille (2020), Goodman-Bacon (2021), Callaway and Sant'Anna (2021), Sun and Abraham (2021), Bugni, Canay, and McBride (2023), Borusyak, Jaravel, and Spiess (2024), Goldsmith-Pinkham, Hull, and Kolesár (2024).

To shed more light on when negative weights are more likely to arise, the paper presents a simple one-factor model in which each component of the summary index is a linear function of a latent scalar factor plus component-specific noise. The model captures the idea that summary index components are noisy proxies for the same latent concept (e.g., women's empowerment). In this model, the **IC** summary index naturally assigns negative weights to components that are *more strongly* associated with the latent factor. Computing the weights for several recently published articles with available data confirms that negative weights arise in real datasets. For example, in Allcott et al. (2020a), who estimate the effect of Facebook deactivation on a summary index of subjective well-being, self-reported happiness receives a weight of $-0.15$; in Carranza et al. (2022), who study skill certification for young job seekers, the labor market summary index assigns a weight of $-1.04$ to being employed and $-3.30$ to weekly earnings. Because the weights determine what the estimated effect on a summary index measures, computing and reporting them is essential for an adequate interpretation of the estimates.

Some empirical articles motivate the use of summary indices by appealing to the improved precision of the resulting estimates, yet provide no formal results.[3] This paper shows that such broad claims about efficiency gains from summary indices appear to lack theoretical support. None of the commonly used summary indices generally minimizes the estimator's asymptotic variance. Tests based on these indices can exhibit high power against certain local alternatives that are unknown ex ante and unlikely to be empirically relevant in every case, but power can be low against alternatives in orthogonal directions. As a result, no summary index uniformly dominates in terms of power. Yet in a simple RCT-like scenario with small, homogeneous treatment effect estimated by difference in means between the outcomes in the treated and the control groups, it is possible to unambiguously rank summary indices by the local asymptotic variance of estimates they produce, with a version of the **IC** summary index achieving the lowest variance. As an alternative to using summary indices, one can analyze the effects on the components of the index separately. However, to ensure correct nominal size, the resulting tests must be adjusted for multiple hypotheses. A simple but conservative way to do so is based on Bonferroni inequalities. The paper shows that tests based on summary indices are not guaranteed to have higher power than such conservative Bonferroni-corrected tests applied to individual components. Concrete examples from real datasets are provided in Section 4.

This paper derives the asymptotic distribution of estimates that use summary indices and proposes two procedures for valid inference: a plug-in estimator of the asymptotic variance, which accounts for the data-dependent weights that default variance estimators in statistical packages

[3]"[**SN** index] gives us **maximal power** to detect an effect on social outcomes, if such an effect is present." Banerjee et al. (2015) (p. 49)

Baranov et al. (2020a) refer to **IC** as "**the most efficient** weighted average of a set of outcomes" (p. 838)

"An index [**SN**] is useful for **increasing power** to the extent that all underlying components move in the same direction...", Bhatt et al. (2024) (p. 20)

"In order to examine heterogeneous effects with **more statistical precision**, column 6 reports on an inverse covariance weighted index [**IC** index] of labor and spending", Kinnan et al. (2024) (p. 279)

"[**SN** index] **improves statistical power** to detect effects that go in the same direction within a domain" Kling, Liebman, and Katz (2007) (p. 89) omit, and a shifted t-test, which tests hypotheses about the component effects without estimating the variance of the weights; the conventional t-test of the null of no effect is a special case of the latter and remains valid. While most of the derivations cover the case of a scalar summary index, the approach extends directly to multiple summary indices constructed from potentially overlapping sets of component outcomes – a common practice in empirical studies – and yields jointly valid confidence intervals for the effects on the indices. Joint inference on several summary indices is currently carried out with the resampling procedure of Westfall and Young (1993) (e.g., Finkelstein et al. (2012), Casey, Glennerster, and Miguel (2012), Haushofer and Shapiro (2016), Blattman, Jamison, and Sheridan (2017), Baranov et al. (2020a), Bessone et al. (2021), Bhatt et al. (2024)), which re-estimates the component and index regressions in each resampling draw; the intervals proposed here require a single estimation.

Overall, individual component analyses seem to be strictly more informative about the studied effects than summary-index-based analysis. Yet **SN** summary indices can provide a reasonable summary of component-level treatment effects, and the estimated effect should be interpreted as the average effect across the components. Beyond that, this paper finds no clear benefit to summary indices, and in particular **IC** summary indices: their weights may obscure the assessment of the treatment, and they offer no statistical power advantage over **SN** summary indices outside the simple RCT-like scenario with the difference-in-means estimator. Nonetheless, should a researcher use summary indices, it appears useful to explicitly report the weights applied to the treatment effects on the summary index's components and to check that these weights are positive. The explicit expressions provided in this paper for **SN** and **IC** summary indices make doing so straightforward. Reporting the weights would make the analysis more transparent and give readers and decision makers more information about the drivers of the reported causal effects.

*Literature* This work relates to the literature on treatment evaluation with multiple outcomes, which dates back to Hotelling (1931), who introduced omnibus tests based on the $T^2$ statistic. Since such tests provide no information about the direction of departures from the null, their usefulness is limited for decision-making, as it is crucial to know whether the treatment benefits or harms recipients. To address this limitation, O'Brien (1984) proposed projection-based tests that target specific one-dimensional alternatives. The two statistics studied in O'Brien (1984) are linear combinations of sample mean differences across outcomes: one a simple average, the other weighted by the inverse variance-covariance matrix – the direct precursors of the **SN** and **IC** summary indices examined in this paper, respectively. These ideas later entered applied economics through (i) Kling, Liebman, and Katz (2007), who introduced the **SN** index to evaluate the Moving to Opportunity program, and (ii) Anderson (2008), who proposed the **IC** index to study the effects of early intervention programs.

A related but distinct literature studies inference with multiple outcomes while retaining component-level hypotheses, typically through adjustments for multiple testing (e.g., Westfall and Young (1993), Romano and Wolf (2005), List, Shaikh, and Xu (2019)). That literature controls error rates across a family of tests; this paper instead studies the estimand, weights, and variance induced by constructing a summary index. Multiple testing becomes relevant below only as a benchmark for componentwise inference.

Independent and concurrent work by Lau (2026) is related, but the two papers differ in scope and exposition. Lau (2026) develops a decision-theoretic approach to choosing index weights, whereas this paper studies the properties of the existing **SN** and **IC** procedures when treatment effects are estimated using the class of IV-like estimators introduced in Section 2. The papers nevertheless obtain some common results, including that **IC** can assign negative weights.

*Notation* Throughout the paper, standard Latin or Greek letters generally denote both vectors and scalars, without relying on boldface. Boldface is reserved for cases where it is important to indicate that an object is a vector or a matrix (rather than a scalar). For any integer $k$, $\vec{1}_k$ denotes the $k$-dimensional vector of ones, and $I_k = (e_{1,k} \cdots e_{k,k})$ is the $k \times k$ identity matrix. All $o(\cdot)$ and $o_p(\cdot)$ terms are to be understood as referring to the limit $n \to \infty$, where, as usual, $n$ denotes the sample size.

## 2. SETUP AND NOTATION

*Data description and summary index construction* The observable data, $\{W_i\}_{i=1}^n := \{\boldsymbol{y}_i, D_i, x_i\}_{i=1}^n$, consist of a vector of outcomes, $\boldsymbol{y}_i \in \mathbb{R}^p$; a cause (treatment, policy intervention, etc.), $D_i \in \mathcal{D} \subseteq \mathbb{R}$, whose effect on $\boldsymbol{y}_i$ is of interest; and additional covariates and instruments, $x_i \in \mathbb{R}^{k-1}$, which, for convenience, include a constant term as their first component. All independent variables are grouped into a single vector, $\tilde{x}_i = (D_i, x_i')' \in \mathbb{R}^k$. Throughout, it is assumed that $0 \in \mathcal{D}$ and that the *control group* is the subsample with $D_i = 0$; its sample moments are used for standardization in the procedures below.

The **SN** summary index, introduced by Kling, Liebman, and Katz (2007), is constructed as follows:

PROCEDURE 1: *Scale Normalized Summary Index.*

1. Switch signs in $\boldsymbol{y}$ if necessary so that the positive direction indicates a "better" outcome;

2. Demean all outcomes in $\boldsymbol{y}$ by subtracting the control-group mean and dividing each outcome by its control-group standard deviation, calling the transformed vector of outcomes $\tilde{\boldsymbol{y}}$;
3. Compute the summary index $\boldsymbol{s} := (s_1, ..., s_n)'$ as:

$$s_i = \left(\vec{1}_p' \, \vec{1}_p\right)^{-1} \vec{1}_p' \, \tilde{\boldsymbol{y}}_i$$

4. Optionally, standardize $\boldsymbol{s}$ once more, i.e., repeat Step 2 for $\boldsymbol{s}$.

The **IC** summary index, introduced by Anderson (2008), is constructed as follows:

PROCEDURE 2: ***Inverse Covariance Summary Index.***

1. Switch signs in $\boldsymbol{y}$ if necessary so that the positive direction indicates a "better" outcome;
2. Demean all outcomes,[4] divide each outcome by its control-group standard deviation, call the transformed vector of outcomes $\tilde{\boldsymbol{y}}$, and compute the variance-covariance matrix of $\tilde{\boldsymbol{y}}$, $\tilde{\boldsymbol{V}}_n := \widehat{\mathrm{Var}}[\tilde{\boldsymbol{y}}]$;
3. Compute the summary index $\boldsymbol{s} := (s_1, ..., s_n)'$ as:

$$s_i = \left(\vec{1}_p' \, \tilde{\boldsymbol{V}}_n^{-1} \, \vec{1}_p\right)^{-1} \vec{1}_p' \, \tilde{\boldsymbol{V}}_n^{-1} \, \tilde{\boldsymbol{y}}_i$$

4. Optionally, standardize $\boldsymbol{s}$ once more, i.e., repeat Step 2 for $\boldsymbol{s}$.

Unlike Procedure 1, Procedure 2 incorporates the inverse variance-covariance matrix of the outcomes. Anderson (2008) refers to this weighting scheme as Generalized Least Squares (GLS) weighting and motivates it as follows: "The GLS weighting procedure [...] increases efficiency by ensuring that outcomes that are highly correlated with each other receive less weight, while outcomes that are uncorrelated and thus represent new information receive more weight." Yet the label "GLS" is somewhat misleading, since the GLS estimator uses the inverse of the *error-term* variance-covariance matrix, not that of the outcomes. However, some empirical works (e.g., Bessone et al. (2021)) do follow the true GLS approach by using the inverse covariance matrix of the estimated residuals. In what follows, I focus on Procedures 1 and 2. Applied implementations differ in details of Steps 2–3: the covariance matrix $\tilde{\boldsymbol{V}}_n$ may be computed on the full sample, as above, on the control group only, or by another sample analogue (see Section 4 for an example), and standardization may use full-sample rather than control-group standard deviations. Such choices alter the weights in Table II only through the sample moments that enter them, and the results of Section 3 apply with the corresponding moments. Applied works also use larger modifications of these procedures: multiple summary indices are constructed from overlapping sets of outcomes or from other summary indices, and other summary index constructions exist based on principal components (Filmer and Pritchett (2001)) or economic weights (Bhatt et al. (2024)). Appendix D provides some additional discussion of how the results apply to these cases.

[4]In some studies, outcomes are demeaned by subtracting the control-group mean as in Procedure **SN**. As shown later, this difference is immaterial for the results (see the discussion in the beginning of Section 3.1).

*Estimand and estimator* I consider the class of estimands that can be expressed as the expected value of $\boldsymbol{y}$ weighted by a mean-zero function of the data, $\omega(\cdot)$. Mogstad, Santos, and Torgovitsky (2018) refer to this class of estimands as *IV-like estimands*.

Definition 1: *IV-like estimand.* An IV-like estimand induced by the weighting function $\omega(\cdot,\cdot)$ and a finite-dimensional parameter $\nu$ has the form:

$$\boldsymbol{\tau} = E[\boldsymbol{y}\, \omega(\tilde{x}, \nu)] \in \mathbb{R}^p, \tag{1}$$

where $\omega(\tilde{x}, \nu)$ is a known (or identified) scalar-valued measurable function satisfying

$$E[\omega(\tilde{x}, \nu)] = 0. \tag{2}$$

As explained in Mogstad, Santos, and Torgovitsky (2018), least-squares (OLS), instrumental variables (IV), and two-stage least squares (TSLS) estimands fit this class; therefore, the setup covers estimates obtained in linear models, linear instrumental-variables models, two-way-fixed-effects models, difference-in-differences models, and (fuzzy) regression-discontinuity designs. Since the notation and dimensions of some variables do not perfectly coincide with those of Mogstad, Santos, and Torgovitsky (2018), I provide a table (Table I) of notable IV-like estimands and their corresponding weighting functions, similar to Table II in Mogstad, Santos, and Torgovitsky (2018).

TABLE I

NOTABLE IV-LIKE ESTIMANDS

| Estimand | $\boldsymbol{\tau}$ | $\omega(\tilde{x},\nu)$ | $\nu$ | Notes |
|---|---|---|---|---|
| Difference in means | $\bar{\boldsymbol{y}}_1 - \bar{\boldsymbol{y}}_0$ | $(\nu(1-\nu))^{-1}(D-\nu)$ | $P(D=1)$ | – |
| OLS (jth components) | $E[\boldsymbol{y}\,\tilde{x}']\,E[\tilde{x}\tilde{x}']^{-1}\,e_{j,k}$ | $\tilde{x}'\,\nu\,e_{j,k}$ | $E[\tilde{x}\tilde{x}']^{-1}$ | – |
| IV (jth components) | $E[\boldsymbol{y}\,\tilde{z}']\,E[\tilde{x}\tilde{z}']^{-1}\,e_{j,k}$ | $\tilde{z}'\,\nu\,e_{j,k}$ | $E[\tilde{x}\tilde{z}']^{-1}$ | $\tilde{z}=(z,1,x')'$, $z$-scalar |
| TSLS (jth components) | $E[\boldsymbol{y}\,\tilde{z}']\,\Pi'\,(\Pi\,E[\tilde{z}\tilde{x}'])^{-1}\,e_{j,k}$ | $\tilde{z}'\,\nu\,e_{j,k}$ | $\Pi'(\Pi\,E[\tilde{z}\tilde{x}'])^{-1}$ | $\Pi = E[\tilde{x}\tilde{z}']\,E[\tilde{z}\tilde{z}']^{-1}$, $z$-vector |

Notes: The table presents some notable IV-like estimands with corresponding weighting functions $\omega()$ and auxiliary parameters $\nu$. Since $\boldsymbol{y}$ is $p$ —dimensional vector, jth components of OLS, IV or TSLS estimates are also $p$-dimensional vectors.

The unknown population expectations and nuisance parameters $(\nu)$ are replaced by their sample analogues and estimates, respectively, giving rise to an estimator that, by analogy, can be referred to as an IV-like estimator.

DEFINITION 2: *IV-like estimator*. An IV-like estimator induced by the weighting function $\omega(\cdot,\cdot)$ and the finite-dimensional parameter $\nu$ has the form:

$$\hat{\boldsymbol{\tau}}_n = \frac{1}{n}\sum_{i=1}^{n} \boldsymbol{y}_i\,\omega(\tilde{x}_i,\hat{\nu}_n), \tag{3}$$

where $\hat{\nu}_n$ is an estimate of $\nu$ satisfying

$$\frac{1}{n}\sum_{i=1}^{n} \omega(\tilde{x}_i,\hat{\nu}_n) = 0. \tag{4}$$

The sample moment condition (4) is the sample analogue of (2) and holds for the estimators in Table I whenever a constant is included among the regressors.

In practice, researchers often apply IV-like estimators to summary indices in place of, or in addition to, applying them to the original outcomes $\boldsymbol{y}$. This yields a scalar estimator:

$$\hat{\beta}_n = \frac{1}{n} \sum_{i=1}^{n} s_{i,n}\, \omega(\tilde{x}_i, \hat{\nu}_n), \tag{5}$$

where $s_{i,n}$ denotes the summary index value of $\boldsymbol{y}_i$. The scalar $\hat{\beta}_n$ is the estimated effect of the treatment on the summary index it is the quantity reported and interpreted in applications. For brevity, I refer to $\hat{\beta}_n$ as the *summary index estimator*.

Since $s_{i,n}$ is an affine function of $\boldsymbol{y}_i$ and (5) is linear in $s_{i,n}$, $\hat{\beta}_n$ is a linear function of $\hat{\boldsymbol{\tau}}_n$ in (3); the next section derives the coefficients of that function.

## 3. THEORETICAL RESULTS

Section 3.1 demonstrates that the summary index estimator $\hat{\beta}_n$ equals a weighted sum of the components of $\hat{\boldsymbol{\tau}}_n$, an IV-like estimator computed with the full outcome vector $\boldsymbol{y}$. The expressions for the weights are provided in Table II. Section 3.2 pertains to large sample statistical properties of $\hat{\beta}_n$, with Section 3.3 focusing on efficiency (statistical power of tests).

### 3.1. INTERPRETATION

PROPOSITION 1 — $\hat{\beta}_n$ as weighted sum: *Let $\hat{\boldsymbol{\tau}}_n$ be an IV-like estimator as in Definition 2, and $\hat{\beta}_n$ be the summary index estimator as in (5), where the summary index is computed by Procedure 1 or 2. Then*

$$\hat{\beta}_n = A_n\, \hat{\boldsymbol{\tau}}_n,$$

*for some row vector of weights $A_n$.*

Thus, if $\hat{\beta}_n$ denotes the estimated treatment effect, then the treatment effect on a summary index equals a weighted sum of the treatment effects on its components.[5] The two-step argument behind this proposition is as follows. First, any summary index produced by Procedures 1 or 2 can be represented as:

$$s_{i,n} = \tilde{A}_n \boldsymbol{y}_i + b_n, \tag{6}$$

for some vector $\tilde{A}_n$ and scalar $b_n$. This follows from the fact that each step in Procedures 1 and 2 – subtraction, scalar or matrix multiplication, and addition – is an affine operation, and the successive application of affine transformations preserves the affine form in (6). Second, the linearity of the estimator (5) and the sample moment condition $\frac{1}{n}\sum_{i=1}^{n} \omega(\tilde{x}_i, \hat{\nu}_n) = 0$ imply:

[5] Kling, Liebman, and Katz (2007) make the same observation but only for the difference-in-means estimator and the **SN** summary index (see their footnote 10, p. 89). In fact, the relation holds much more broadly: it applies to the entire class of IV-like estimators and to any Summary index that can be represented as an affine transformation.

$$\hat{\beta}_n = \frac{1}{n}\left(\tilde{A}_n \sum_{i=1}^{n} \boldsymbol{y}_i \omega(\tilde{x}, \hat{\nu}_n) + b_n \sum_{i=1}^{n} \omega(\tilde{x}, \hat{\nu}_n)\right) = \tilde{A}_n \hat{\boldsymbol{\tau}}_n,$$

hence the proposition, with $A_n = \tilde{A}_n$.

Two consequences of this result are worth noting. First, it is immaterial whether outcomes are demeaned and, if they are, whether demeaning is done using the control-group mean or the full-sample mean. Any constant $b_n$ cancels out by the properties of the estimator. Thus, the demeaning step in Procedures 1 and 2 can be skipped, as it does not affect the estimate. Second, if some components of $\boldsymbol{y}$ are missing for some observations, the common practice is to construct the summary index from the components observed for each observation, renormalizing the weights over them. The weights then differ across observations according to their missingness patterns, and no single vector $A_n$ satisfies Proposition 1: the estimated effect on the summary index is no longer a fixed linear combination of the component-level estimates. The alternative that preserves the representation if data are missing is to estimate the component effects under explicit missing-data assumptions and combine them with fixed weights, $A_n \hat{\boldsymbol{\tau}}_n$; Appendix E gives some additional details.

Thus, estimating the treatment effect on a summary index provides a simple yet indirect way to summarize the vector of component-level treatment effects in the form of their weighted sum. I next analyze the weights. Let $\hat{S}_0$ be a $p \times p$ diagonal matrix with the control-group standard deviations of $\boldsymbol{y}$ on its main diagonal. Using this matrix, define the vector of control-group-standardized treatment effects:

$$\hat{\boldsymbol{\tau}}_{n,\,\mathrm{st}} := \hat{S}_0^{-1}\, \hat{\boldsymbol{\tau}}_n. \tag{7}$$

The transformation in (7) expresses each treatment effect in units of the control-group standard deviation of its component, so that the components are comparable and a weighted average of them does not depend on the units in which the components are measured. When all components are measured on the same scale – e.g., the transportation availability summary index in Asher and Novosad (2020) aggregates binary indicators of the availability of different modes of transportation – this purpose does not apply.

Define the arithmetic average of the control-group-standardized component-level treatment-effect estimates as

$$\overline{\tau}_{n,\,\mathrm{st}} := p^{-1}\, \vec{1}_p'\, \hat{\boldsymbol{\tau}}_{n,\,\mathrm{st}}.$$

Proposition 2 — Weights of summary index: *Let $\hat{\boldsymbol{\tau}}_{n,\,\mathrm{st}}$ be the vector of control-group-standardized IV-like estimates and $\hat{\beta}_n$ be the summary index estimate. Then*

$(i)$ *For* **SN** *summary index without final standardization* $($*step 4 of Procedure 1*$)$*:*

$$\hat{\beta}_n = \overline{\tau}_{n,\,\mathrm{st}}$$

*Thus, the summary index estimator is the arithmetic mean of the control-group-standardized component treatment effects.*

$(ii)$ *For* **SN** *summary index with final standardization:*

$$\hat{\beta}_n = c \times \overline{\tau}_{n,\,\mathrm{st}},$$

*for* $c = p/\sqrt{\vec{1}_p'\hat{R}_0\vec{1}_p} \geq 1$*, where* $\hat{R}_0$ *is the control-group correlation matrix of the components. Thus, the summary index estimator is a multiple of the arithmetic mean of the control-group-standardized component treatment effects.*

$(iii)$ *For* **IC** *summary index:*

$$\hat{\beta}_n = A_n^{\mathbf{IC}}\, \hat{\boldsymbol{\tau}}_{n,\,\mathrm{st}},$$

*where the entries of* $A_n^{\mathbf{IC}}$ *can exceed 1 in magnitude or be negative. Without final standardization* $($*step 4 of Procedure 2*$)$*, the weights sum to one,* $A_n^{\mathbf{IC}}\,\vec{1}_p = 1$*; with standardization, this equality generally fails.*

The argument behind this proposition is as follows. Table II below provides expressions for the weights induced by the summary index procedures; they are derived in Proposition A.1 in Appendix A and imply points (i) and (ii) as well as the representation and the sum-to-one statement in point (iii). The claim in point (iii) that **IC** weights can exceed 1 in magnitude or be negative is an existence statement; Section 4 documents both cases in data from recently published articles, and Proposition 3 gives conditions under which negative weights arise.

Note that different summary index constructions generally target different scalar summaries of the component-level treatment effects. These targets can differ even when all components of $\hat{\boldsymbol{\tau}}_{n,\,\mathrm{st}}$ are the same, because the corresponding weights need not sum to one (or any other pre-specified constant).

As Proposition 2 implies, only the **SN** summary index without final standardization yields an estimate that is a summary of the control-group-standardized component effects alone, namely their arithmetic mean. If the index is standardized, the reported effect equals that arithmetic mean multiplied by $c = p/\sqrt{\vec{1}_p'\hat{R}_0\vec{1}_p} \geq 1$, where $\hat{R}_0$ is the control-group correlation matrix of the components: the factor depends only on the correlations among the components, not on the treatment effects, and exceeds one unless the components are perfectly correlated. For an **IC** summary index, negative weights can give the estimate $\hat{\beta}_n$ the opposite sign to every component of $\hat{\boldsymbol{\tau}}_{n,\,\mathrm{st}}$ when these share the same sign. Thus, if an **IC** summary index is used, the weights the procedure assigns to the components, given in Table II, should be computed and reported; if any weight is negative, the estimate is not a summary of the component effects.

TABLE II

WEIGHTS ON CONTROL-GROUP-STANDARDIZED TREATMENT EFFECTS

| Summary index | Weight vector |
|---|---|
| **SN** no final standardization | $p^{-1}\,\vec{1}_p'$ |
| **SN** with final standardization | $\left(\vec{1}_p'\,\hat{S}_0^{-1}\,\hat{\Sigma}_0\,\hat{S}_0^{-1}\,\vec{1}_p\right)^{-\frac{1}{2}}\,\vec{1}_p'$ |
| **IC** no final standardization | $\left(\vec{1}_p'\,\hat{S}_0\,\hat{\Sigma}^{-1}\,\hat{S}_0\,\vec{1}_p\right)^{-1}\,\vec{1}_p'\,\hat{S}_0\,\hat{\Sigma}^{-1}\hat{S}_0$ |
| **IC** with final standardization | $\left(\vec{1}_p'\,\hat{S}_0\,\hat{\Sigma}^{-1}\,\hat{\Sigma}_0\,\hat{\Sigma}^{-1}\,\hat{S}_0\,\vec{1}_p\right)^{-\frac{1}{2}}\,\vec{1}_p'\,\hat{S}_0\,\hat{\Sigma}^{-1}\hat{S}_0$ |

Notes: Each row reports the row vector $w_n$ such that $\hat{\beta}_n = w_n\,\hat{\boldsymbol{\tau}}_{n,\,\mathrm{st}}$, where $\hat{\boldsymbol{\tau}}_{n,\,\mathrm{st}} = \hat{S}_0^{-1}\hat{\boldsymbol{\tau}}_n$ are the component-level estimates measured in control-group standard deviations and $\hat{S}_0$ is the diagonal matrix of control-group standard deviations of $\boldsymbol{y}$. In terms of Proposition 1, $w_n = A_n\hat{S}_0$. The matrices $\hat{\Sigma}$ and $\hat{\Sigma}_0$ are the sample covariance matrices of $\boldsymbol{y}$ in the full sample and in the control group, respectively;

Table III illustrates the four weight vectors for a five-component labor market summary index from Carranza et al. (2022), computed from the replication data (the application is described in Section 4).

TABLE III

LABOR MARKET INDEX: COMPONENTS AND WEIGHTS

| Component | **SN** no std | **SN** std | **IC** no std | **IC** std |
|---|---|---|---|---|
| Employed | 0.20 | 0.23 | **−0.81** | **−1.04** |
| Hours worked | 0.20 | 0.23 | 1.84 | 2.36 |
| Earnings (weekly) | 0.20 | 0.23 | **−2.58** | **−3.30** |
| Wage (hourly) | 0.20 | 0.23 | 2.29 | 2.93 |
| Written contract | 0.20 | 0.23 | 0.26 | 0.33 |
| **Total sum** | 1.00 | **1.14** | 1.00 | **1.28** |

Notes: The table lists the components of the "Labor market summary index" defined in Carranza et al. (2022) and, for each of the four summary-index procedures studied, the implicit component weights computed as in Table II. The final row reports the total weight for each summary index. Boldface indicates negative weights and total weights greater than 1.

#### 3.1.1. WHEN ARE WEIGHTS NEGATIVE?

Summary indices can be motivated by the idea that several observed outcomes are noisy proxies for the same latent concept. The following single-factor measurement model formalizes that idea:

$$\boldsymbol{y} = \lambda F + \varepsilon, \varepsilon \sim (0, \Psi), \tag{8}$$

where $F$ is an unobserved, latent scalar factor, $\lambda$ is a $p$-dimensional vector of factor loadings, so $\lambda_j$ is the response of component $j$ to a unit change in the latent quantity $F$.

ASSUMPTION 1: .

1. Errors are uncorrelated: $\Psi = \text{Diag}(\sigma_1^2, ..., \sigma_p^2)$.
2. The factor and the errors are uncorrelated in the full sample and in the control group: $\text{Cov}(F, \varepsilon) = 0$ and $\text{Cov}(F, \varepsilon \mid D = 0) = 0$.
3. The treatment does not affect the error variance: $\text{Var}[\varepsilon \mid D = 0] = \Psi$. The distribution of $F$ is unrestricted and may depend on $D$ and $\tilde{x}$.
4. Components of $\boldsymbol{y}$ are non-negatively associated with the factor: $\lambda_j \geq 0$ for all $j$, and $\lambda \neq 0$.

This is a simplified version of the model studied by Jöreskog and Goldberger (1975).

Proposition 3 — Negative weights in single-factor model: *Consider the single-factor model (8) under Assumption 1. Let* $\mathrm{Var}_0(y_j)$ *denote the control-group variance of* $y_j$*, and* $A^{\mathrm{IC}} = (a_1, ..., a_p)$ *be the population version of a vector of weights corresponding to the* ***IC*** *summary index. Then there exists a threshold* $r > 0$ *such that*

$$a_j < 0 \Leftrightarrow \frac{\lambda_j}{\sqrt{\mathrm{Var}_0(y_j)}} > r.$$

In the single-factor model, therefore, a component receives a negative weight in the **IC** summary index if and only if its loading relative to its control-group standard deviation exceeds the threshold $r$; the components at risk are those with the largest such ratios. The single-factor structure is not necessary for negative weights, however: they arise outside (8) as well, so whether an **IC** summary index assigns negative weights in a given application can only be established by computing the weights in Table II.

The argument behind Proposition 3 applies the Sherman–Morrison formula to $\Sigma^{-1}$ in the expression for the weights in Table II; the details are in Appendix A.

### 3.2. LARGE SAMPLE STATISTICAL PROPERTIES OF ESTIMATES

This section derives the large sample properties of $\hat{\beta}_n$ for every IV-like estimator of Definition 2. The only requirement on $\hat{\boldsymbol{\tau}}_n$ is an asymptotically linear representation, which these estimators admit under standard regularity conditions.

Assumption 2: *Asymptotic linearity of estimates.*

$$\hat{\boldsymbol{\tau}}_n = \boldsymbol{\tau} + \frac{1}{n}\sum_{i=1}^{n} \vartheta(W_i) + o_p\left(\frac{1}{\sqrt{n}}\right), \tag{9}$$

for a measurable *influence function* $\vartheta(\cdot)$ such that $E[\vartheta(W)] = 0, E[\|\vartheta(W)\|^2] < \infty$.

All notable IV-like estimators generally satisfy this assumption, and expressions for their influence functions $\vartheta$ are known.

Example 1: *OLS Estimate*. For the linear projection:

$$\boldsymbol{y} = \boldsymbol{\tau} D + \Gamma x + u, E[u\,\tilde{x}'] = 0, E[\tilde{x}\tilde{x}']^{-1} = \nu < \infty,$$

and OLS estimate:

$$\hat{\boldsymbol{\tau}}_n = \frac{1}{n}\sum_{i=1}^{n} \boldsymbol{y}_i\, \tilde{x}_i' \underbrace{\left(\frac{1}{n}\sum_{i=1}^{n} \tilde{x}_i \tilde{x}_i'\right)^{-1}}_{:=\hat{\nu}} e_{1,k} =$$

$$= \boldsymbol{\tau} + \frac{1}{n}\sum_{i=1}^{n} u_i\, \tilde{x}_i'\, \nu\, e_{1,k} + \frac{1}{n}\sum_{i=1}^{n} u_i\, \tilde{x}_i'\, (\hat{\nu}_n - \nu)\, e_{1,k}, \tag{10}$$

so the influence function in the decomposition (9) equals:

$$\vartheta(W; \boldsymbol{\tau}, \Gamma, \nu) = u\, \tilde{x}'\, \nu\, e_{1,k} = (\boldsymbol{y} - [\boldsymbol{\tau}, \Gamma]\, \tilde{x})\, \tilde{x}'\, \nu\, e_{1,k}, \tag{11}$$

where, as before, $e_{1,k} = (1, 0, ..., 0)'$. The last term in (10) is $o_p\left(\frac{1}{\sqrt{n}}\right)$ under additional standard least-squares moment conditions.

The representation in Assumption 2 can be obtained generically by computing the Hadamard derivative of an IV-like estimand with respect to the data distribution. This approach has been used to show that the weights that summary indices assign to the components admit a similar representation.

Assumption 3: *Asymptotic linearity of weights.*

$$\hat{A}_n = A + \frac{1}{n}\sum_{i=1}^{n} \psi(W_i) + o_p\left(\frac{1}{\sqrt{n}}\right), \tag{12}$$

for some non-stochastic $A$ and a measurable function $\psi(\cdot)$ such that $E[\psi(W)] = 0, E[\|\psi(W)\|^2] < \infty$.

Proposition A.2 in Appendix A provides an expression for the influence function $\psi$ for the non-standardized **SN** summary index – the only procedure studied here whose weights are non-negative and sum to one in general

Proposition 4 — Large sample properties of $\hat{\beta}_n$: *Under Assumptions 2 and 3, $\hat{\beta}_n$ is consistent for $\beta := A\boldsymbol{\tau}$,*

$$\hat{\beta}_n \xrightarrow{p} \beta \;\; as \;\; n \to \infty,$$

*and asymptotically normal,*

$$\sqrt{n}\left(\hat{\beta}_n - \beta\right) \xrightarrow{d} \mathcal{N}(0, \varsigma) \;\; as \;\; n \to \infty, \tag{13}$$

*where*

$$\varsigma = \text{Var}(A\,\vartheta(W) + \psi(W)\,\boldsymbol{\tau}), \tag{14}$$

*and* $\vartheta(W) \in \mathbb{R}^{p \times 1}$, $\psi(W) \in \mathbb{R}^{1 \times p}$ *are the influence functions defined in Assumptions 2 and 3.*

The proposition follows from taking the product of the two representations in Assumptions 2 and 3:

$$\hat{\beta}_n = \underbrace{A\boldsymbol{\tau}}_{=\beta} + \frac{1}{n}\sum_{i=1}^{n}(A\,\vartheta(W_i) + \psi(W_i)\,\boldsymbol{\tau}) + o_p\left(\frac{1}{\sqrt{n}}\right),$$

and then applying the Lindeberg-Levy central limit theorem. A short proof is included in Appendix A for completeness.

Note that when summary indices are used as left-hand-side variables in regressions, commonly used statistical packages produce standard errors that generally fail to estimate the true variability of $\hat{\beta}_n$ as given by the asymptotic variance expression in (14).

Example 1 (continued): *OLS Estimate*. Let $\left(\hat{\beta}_n, \hat{\Lambda}_n\right)$ be OLS estimates from:

$$s = \hat{\beta}_n\, D + \hat{\Lambda}_n x + \hat{\varepsilon}, \tag{15}$$

and $\left(\hat{\boldsymbol{\tau}}_n, \hat{\Gamma}_n\right)$ be OLS estimates from:

$$\boldsymbol{y} = \hat{\boldsymbol{\tau}}_n\, D + \hat{\Gamma}_n x + \hat{u}.$$

One can show that $\hat{\varepsilon}_i = A_n \hat{u}_i$, $\forall i$. The reported sandwich variance estimate from regression (15) (for example, the HC1 version) then equals:

$$\widehat{\mathrm{Var}}\left(\hat{\beta}_n\right) = e'_{1,k}\left(\left(\frac{1}{n}\sum_{i=1}^{n}\tilde{x}_i\tilde{x}'_i\right)^{-1}\left(\frac{1}{n}\sum_{i=1}^{n}\tilde{x}_i\tilde{x}'_i\hat{\varepsilon}_i^2\right)\left(\frac{1}{n}\sum_{i=1}^{n}\tilde{x}_i\tilde{x}'_i\right)^{-1}\right)e_{1,k} =$$

$$= \frac{1}{n}\sum_{i=1}^{n}\hat{\varepsilon}_i\left(e'_{1,k}\,\hat{\nu}_n\,\tilde{x}_i\right)\left(\tilde{x}'_i\,\hat{\nu}_n\,e_{1,k}\right)\hat{\varepsilon}_i = A_n\left(\frac{1}{n}\sum_{i=1}^{n}\hat{u}_i\left(e'_{1,k}\hat{\nu}_n\tilde{x}_i\right)\left(\tilde{x}_i\hat{\nu}_n e_{1,k}\right)\hat{u}'_i\right)A'_n,$$

which, because of (11), at best estimates $\mathrm{Var}(A\,\vartheta(W))$ and ignores the second term in (14), arising from the variability of the weights. Generally this leads to invalid confidence intervals.

The example illustrates a general point: because conventional packages mechanically ignore the variability in summary index weights, the confidence intervals they report for $\beta$ are generally invalid. The resulting distortion of confidence intervals may nevertheless be immaterial for conclusions about statistical significance when the standardized component treatment effects, $\boldsymbol{\tau}_{st} := S_0^{-1}\,\boldsymbol{\tau}$, are small. The term omitted by conventional variance estimators is $\Delta_1 + \Delta_2$, where $\Delta_1 := 2\,\mathrm{Cov}(A\,\vartheta(W), \psi(W)\,\boldsymbol{\tau})$ and $\Delta_2 := \mathrm{Var}(\psi(W)\,\boldsymbol{\tau})$; Proposition A.3 in Appendix A gives their expressions for the **SN** summary index. $\Delta_1$ is linear and $\Delta_2$ quadratic in $\boldsymbol{\tau}_{st}$, so the omitted term is $O(\|\boldsymbol{\tau}_{st}\|)$ as $\boldsymbol{\tau}_{st} \to 0$ and vanishes under $\boldsymbol{\tau} = 0$.

Valid inference can be implemented in two ways: by estimating the asymptotic variance in (14), which yields confidence intervals for $\beta$, or by shifting the estimator, which yields tests of hypotheses about $\boldsymbol{\tau}$ without estimating the variance of the weights. The first requires the two influence functions in (14). For commonly used IV-like estimators, $\vartheta$ is known. Once $\psi$ is available – Proposition A.2 in Appendix A gives it for **SN** summary indices without final standardization – one can replace the population quantities in $A\,\vartheta(W_i) + \psi(W_i)\,\boldsymbol{\tau}$ with sample analogues. Under conditions stated in Appendix B, this plug-in procedure yields asymptotically valid standard errors and confidence intervals.

The second approach starts from the observation that the t-test reported in the summary index regression has the correct asymptotic size for the null hypothesis $\boldsymbol{\tau} = 0$: under this null the omitted variance term disappears, so conventional variance estimates are correct. This extends to any fixed vector $\boldsymbol{\tau}_0$ by shifting the estimator: $\hat{A}_n(\hat{\boldsymbol{\tau}}_n - \boldsymbol{\tau}_0) = \hat{\beta}_n - \hat{A}_n\boldsymbol{\tau}_0$. Writing $\hat{A}_n(\hat{\boldsymbol{\tau}}_n - \boldsymbol{\tau}_0) = A(\hat{\boldsymbol{\tau}}_n - \boldsymbol{\tau}_0) + \left(\hat{A}_n - A\right)(\hat{\boldsymbol{\tau}}_n - \boldsymbol{\tau}_0)$, the second term is $o_p\left(n^{-\frac{1}{2}}\right)$ under $\boldsymbol{\tau} = \boldsymbol{\tau}_0$ whenever the weights are consistent, so the null distribution of the shifted statistic is that of $A(\hat{\boldsymbol{\tau}}_n - \boldsymbol{\tau})$, whose variance is the quantity conventional variance estimators target. The null variance can therefore be estimated by applying conventional variance estimators to the shifted statistic, without estimating $\psi$ and without Assumption 3. The following result summarizes these considerations; its proof is in Appendix A.

Proposition 5 — t-test for shifted estimator: *Suppose Assumption 2 holds, $\hat{A}_n \xrightarrow{p} A$, and let $\hat{V}_n$ be a variance estimator with $\hat{V}_n \xrightarrow{p} \operatorname{Var}(A\,\vartheta(W)) > 0$, such as the sandwich estimator from the summary index regression in Example 1. For a fixed $\boldsymbol{\tau}_0 \in \mathbb{R}^p$, define*

$$T_n(\boldsymbol{\tau}_0) := \frac{\sqrt{n}\hat{A}_n(\hat{\boldsymbol{\tau}}_n - \boldsymbol{\tau}_0)}{\sqrt{\hat{V}_n}} = \frac{\sqrt{n}\big(\hat{\beta}_n - \hat{A}_n\boldsymbol{\tau}_0\big)}{\sqrt{\hat{V}_n}}. \tag{16}$$

*Then, under $H_0 : \boldsymbol{\tau} = \boldsymbol{\tau}_0$, $T_n(\boldsymbol{\tau}_0) \xrightarrow{d} \mathcal{N}(0,1)$, so the test that rejects when $|T_n(\boldsymbol{\tau}_0)| > z_{1-\frac{\alpha}{2}}$ has asymptotic size $\alpha$. In particular, $T_n(\mathbf{0})$ is the conventional t-statistic reported for the summary index regression, which is therefore an asymptotically valid test statistics for $H_0 : \boldsymbol{\tau} = \mathbf{0}$.*

Proposition 5 requires only consistency of the weights, so it applies to every summary index studied here, including **IC**, without deriving $\psi$. Two qualifications matter in practice. First, the null hypothesis is the vector hypothesis $\boldsymbol{\tau} = \boldsymbol{\tau}_0$, not the scalar hypothesis $\beta = A\boldsymbol{\tau}_0$: if $\boldsymbol{\tau} \neq \boldsymbol{\tau}_0$ but $A\boldsymbol{\tau} = A\boldsymbol{\tau}_0$, the term $\psi(W)(\boldsymbol{\tau} - \boldsymbol{\tau}_0)$ does not vanish and $T_n(\boldsymbol{\tau}_0)$ has asymptotic variance $\operatorname{Var}(A\vartheta(W) + \psi(W)(\boldsymbol{\tau} - \boldsymbol{\tau}_0))/\operatorname{Var}(A\vartheta(W)) \neq 1$ in general, so the conventional t-test does not control size for $H_0 : \beta = \beta_0$. Second, for the same reason, inverting (16) yields a confidence region for $\boldsymbol{\tau}$, not a confidence interval for $\beta$; the latter requires the corrected variance in (14).

### 3.3. EFFICIENCY

Several empirical studies motivate summary indices by appealing to gains in efficiency or statistical power (see e.g., Footnote 3). The results above show that such claims do not hold in general. In (14), the asymptotic variance of $\hat{\beta}_n$ equals $\operatorname{Var}(A\vartheta(W) + \psi(W)\boldsymbol{\tau})$; hence it depends on the limiting weights $A$, the influence functions $\vartheta$ and $\psi$ of the component estimator and of the estimated weights, and the true treatment-effect vector $\boldsymbol{\tau}$. Therefore, neither **SN** nor **IC** can be uniformly most efficient without further restrictions. In what follows, I proceed by imposing such restrictions sequentially. Under local alternatives (Assumption 4 in Section 3.3.1), the contribution of the estimated weights to the asymptotic variance vanishes, and it becomes straightforward to characterize the treatment-effect directions against which each summary index has high local power. There is still no uniform ranking across directions. Under stronger restrictions (Assumption 5 in Section 3.3.2): a randomized experiment, difference-in-means estimation, homoskedastic component outcomes, and equal component effects in control-group-standard-deviation units, the **IC** summary index achieves the largest local power among all weighting procedures.

An alternative to constructing an index is to test the component hypotheses individually with a multiplicity correction. The local power of the two-sided Bonferroni rule is given in Proposition C.1 in Appendix C; it is not generally below that of index-based tests, and Section 4 gives configurations in which it is higher.

### 3.3.1. LOCAL POWER AND OPTIMAL DIRECTIONS

Consider local alternatives $\boldsymbol{\tau}_n = \frac{\boldsymbol{\delta}}{\sqrt{n}}$. This sequence represents alternatives whose magnitude shrinks at the same rate as the sampling error of $\hat{\boldsymbol{\tau}}_n$ and is commonly used to compare the asymptotic power of tests (see e.g., van der Vaart (2000)).

ASSUMPTION 4: *Local DGP.* Consider the sequence of local alternatives $\boldsymbol{\tau}_n = \frac{\boldsymbol{\delta}}{\sqrt{n}}$, so that the estimator admits the representation

$$\hat{\boldsymbol{\tau}}_n = \frac{\boldsymbol{\delta}}{\sqrt{n}} + \frac{1}{n}\sum_{i=1}^{n} \vartheta(W_i) + o_p\left(\frac{1}{\sqrt{n}}\right), \tag{17}$$

where, under the $n$th local DGP, $E_n[\vartheta(W)] = 0$, $\mathrm{Var}_n[\vartheta(W)] = V$ – a positive definite matrix – and $\boldsymbol{\delta} \in \mathbb{R}^p \setminus \{\mathbf{0}\}$ is the local alternative direction.

The scaling $\boldsymbol{\tau}_n = \frac{\boldsymbol{\delta}}{\sqrt{n}}$ captures the idea that $\boldsymbol{\tau}$ is small in magnitude relative to the sample size. For a sequence of tests $\varphi_n$, the asymptotic local power is the limit, assuming it exists, along such a sequence of local alternatives of the probability of rejecting the null hypothesis,

$$\pi := \lim_{n\to\infty} P_n(\varphi_n = 1).$$

PROPOSITION 6 — Local power and optimal direction: *Under Assumption 3 and Assumption 4, for a scalar summary index with nonzero population weighting row vector $A$, let $\beta_n := A\boldsymbol{\tau}_n$ denote the local sequence of the population treatment effect on the summary index. Then*

$$\sqrt{n}\big(\hat{\beta}_n - \beta_n\big) \xrightarrow{d} \mathcal{N}(0, AVA'). \tag{18}$$

$(i)$ **(*Local power*)** *For the asymptotic-variance-normalized statistic*

$$T_n^{\mathrm{AV}} := \frac{\sqrt{n}\hat{\beta}_n}{\sqrt{AVA'}}$$

*and the two-sided test*

$$\varphi_n^{\mathrm{AV}} = 1\Big\{|T_n^{\mathrm{AV}}| > z_{1-\alpha/2}\Big\}, \tag{19}$$

*the asymptotic local power equals*

$$\pi(\Delta) := \Phi\Big(\Delta - z_{1-\alpha/2}\Big) + \Phi\Big(-\Delta - z_{1-\alpha/2}\Big), \tag{20}$$

*where*

$$\Delta := \frac{A\boldsymbol{\delta}}{\sqrt{AVA'}}.$$

$(ii)$ **(*Optimal direction*)** *Among local alternatives with $\boldsymbol{\delta}'V^{-1}\boldsymbol{\delta}$ fixed, the local power is maximized at*

$$\boldsymbol{\delta}_A^* \propto VA'. \tag{21}$$

(18) follows because, under Assumption 4,

$$\sqrt{n}\Big(\hat{\beta}_n - \beta_n\Big) = \frac{1}{\sqrt{n}}\sum_{i=1}^{n} A\,\vartheta(W_i) + o_p(1),$$

so the stochastic weights do not contribute to the asymptotic variance. Since $\sqrt{n}\beta_n = A\boldsymbol{\delta}$, (18) implies $T_n^{\mathrm{AV}} \xrightarrow{d} \mathcal{N}(\Delta, 1)$, which gives the local power statement. The local power increases in $|\Delta|$, which is maximized at $\boldsymbol{\delta}_A \propto VA'$. The normalization of $\boldsymbol{\delta}$ in (ii) fixes $\boldsymbol{\delta}'V^{-1}\boldsymbol{\delta}$, the noncentrality parameter of the Wald test of $\boldsymbol{\tau} = \mathbf{0}$ based on all $p$ components, so that directions are compared among alternatives that the joint test detects equally well. Another reason for this normalization is that $\boldsymbol{\delta}'V^{-1}\boldsymbol{\delta}$ is invariant to the units in which the components are measured, unlike the Euclidean length $\boldsymbol{\delta}'\boldsymbol{\delta}$. A formal proof is in Appendix A.

Corollary 3.1 — Optimal directions for **SN** and **IC**: *Suppose that, along the sequence of DGPs satisfying Assumption 4, the sample covariance matrix of $\boldsymbol{y}$, $\hat{\Sigma}$, and the control-* *group covariance matrix, $\hat{\Sigma}_0$, satisfy $\hat{\Sigma} \xrightarrow{p} \Sigma$ and $\hat{\Sigma}_0 \xrightarrow{p} \Sigma_0$ as $n \to \infty$, where $\Sigma$ and $\Sigma_0$ are positive definite, non-stochastic matrices. Under Assumption 3, the optimal local directions in (21) for **SN** and **IC** summary indices are*

$$\boldsymbol{\delta}^*_{\mathbf{SN}} \propto V\, S_0^{-1}\, \vec{1}, \quad \boldsymbol{\delta}^*_{\mathbf{IC}} \propto V\, \Sigma^{-1}\, S_0\, \vec{1},$$

*where* $S_0 = \operatorname{Diag}(\Sigma_0)^{\frac{1}{2}}$.

That is, the local-power-maximizing direction for the test (19) based on the **SN** (respectively the **IC**) summary index is proportional to $\boldsymbol{\delta}^*_{\mathbf{SN}}$ (respectively $\boldsymbol{\delta}^*_{\mathbf{IC}}$). Both directions depend on unknown quantities and represent somewhat arbitrary directions in the space of possible treatment-effect directions. In one simple scenario, described in the following subsection, the results can be refined further, and the summary indices studied here can be ranked in terms of the test's local power.

#### 3.3.2. WHEN **IC** WEIGHTING IS OPTIMAL

ASSUMPTION 5: *RCT with homoskedasticity, small effect, and difference-in-means.* Let $(D_i, \boldsymbol{u}_i)$ be i.i.d., with $D_i \in \{0,1\}$, $\nu := P(D_i = 1) \in (0,1)$, $E[\boldsymbol{u}_i \mid D_i] = \mathbf{0}$, and $E[\boldsymbol{u}_i \boldsymbol{u}_i' \mid D_i] = \mathcal{V}$, where $\mathcal{V}$ is positive definite. The local outcomes satisfy

$$\boldsymbol{y}_{i,n} = \alpha + \frac{\boldsymbol{\delta}}{\sqrt{n}} D_i + \boldsymbol{u}_i, i = 1, ..., n.$$

The component-level treatment-effect estimator is the difference in means,

$$\hat{\boldsymbol{\tau}}_n = \bar{\boldsymbol{y}}_1 - \bar{\boldsymbol{y}}_0.$$

The sequence of data-generating processes satisfying Assumption 5 can describe a randomized controlled trial (RCT) in which treatment $D$ is randomly assigned, the variances of the error term are the same in the treated and control groups, the treatment effect is local, and component effects are estimated by difference in means. The homoskedasticity restriction in Assumption 5 is closely related to treatment-effect homogeneity. In the potential-outcome representation

$$\boldsymbol{y}_i = \boldsymbol{y}_i(0) + D_i(\boldsymbol{y}_i(1) - \boldsymbol{y}_i(0)),$$

the error term can be written as

$$\boldsymbol{u}_i = (\boldsymbol{y}_i(0) - \alpha) + D_i\left((\boldsymbol{y}_i(1) - \boldsymbol{y}_i(0)) - \frac{\boldsymbol{\delta}}{\sqrt{n}}\right). \tag{22}$$

Thus, homogeneous component-level treatment effects imply homoskedasticity across treatment arms.

PROPOSITION 7 — Local power optimality of **IC**: *Suppose Assumption 5 holds and $\boldsymbol{\delta} = c\, S_0\, \vec{1}_p$ for some $c \neq 0$. Over all nonzero row vectors $A$, the asymptotic local power (20) of the test (19) is maximized at*

$$A \propto \vec{1}_p'\, S_0\, \mathcal{V}^{-1}.$$

*This is the limiting **IC** weighting vector.*

This clarifies the limits of the efficiency claims quoted in Footnote 3. The result above shows that the power advantage of the **IC** summary index is theoretically justified only in a narrow case, which is unlikely to describe many empirical applications.

## 4. EMPIRICAL ILLUSTRATIONS

This section presents three recently published empirical studies that use summary indices. The empirical exercises use the publicly available data and code deposits associated with these studies: Allcott et al. (2020b), Baranov et al. (2020b), and Carranza et al. (2023). The goal is threefold: first, to illustrate how summary indices are constructed and used in practice; second, to exploit the availability of data in all three cases to compute the implicit weights that summary indices assign to the components (Table II), demonstrating that the problems outlined in Proposition 2 are not merely theoretical possibilities but can arise in real datasets; third, for applications in which component-level estimates and covariance matrices can be reconstructed, to evaluate the local-power formulas derived in Section 3.3 and compare summary-index-based tests with Bonferroni-corrected component tests.

The section reports weights and local powers but not the estimated effects on the indices: by Proposition 1, each summary index estimate is a weighted sum of the component estimates with its own weights, so **SN** and **IC** estimate different parameters, and comparing their values is uninformative. Which weighted sum of the component treatment effects a summary index estimates is determined by its weights, so the section reports and compares the weights instead.

### 4.1. EXAMPLE A: EFFECT OF FACEBOOK DEACTIVATION (ALLCOTT ET AL. (2020A))

The article studies the effects of Facebook on (i) time allocation and substitution toward other online and offline activities, (ii) news knowledge and political polarization, (iii) subjective well-being, and (iv) post-experiment demand for and valuation of Facebook. It employs a variant of the **IC** summary index with standardization to quantify news knowledge, political polarization, and subjective well-being.[6] The paper's main conclusions–that Facebook deactivation causally reduces political polarization and increases subjective well-being–were supported by linear instrumental-variable estimates in which random assignment of the incentive to deactivate a Facebook account is used as an instrument for the endogenous decision to deactivate, and the aforementioned indices are used as outcomes.

The subjective well-being index combines survey-based measures of happiness, life satisfaction, loneliness, depression, anxiety, absorption, and boredom over the previous four weeks. It also includes momentary measures collected through daily text messages: current happiness, positive emotion, and the absence of loneliness. Table IV reports these components and the corresponding weights that different summary index procedures assign to the component-level treatment effects, computed using the formulas in Table II. The last row of Table IV presents the sum of the weights. As the table shows, "happiness" receives a negative weight in the estimated treatment effect on the **IC** summary index of well-being. In addition, when the construction of the summary index includes final standardization, the weights sum to more than one.

The political polarization index combines measures of affective polarization toward the two major parties and the president, anger toward the opposing party, and exposure to news congenial to one's own party. It also includes measures of polarization in political opinions, beliefs about current events, and voting preferences. Table V reports the weights assigned to the components of this summary index. Similar problems are evident: the **IC** summary index assigns negative weights to some components, while procedures involving final standardization produce weights that sum to more than one.

[6]Instead of the covariance matrix, when constructing the Summary index, the paper computes the average inner product of control-group-standardized outcome vectors, but averaged over the entire sample (thus it is neither the full-sample nor the control-group covariance matrix), so the weights do not perfectly match those obtained from Procedure 2, yet the differences are immaterial.

TABLE IV

WELL-BEING SUMMARY INDEX: COMPONENTS AND WEIGHTS

| Component | **SN** no std | **SN** std | **IC** no std | **IC** std |
|---|---|---|---|---|
| Happiness | 0.10 | 0.14 | **−0.10** | **−0.15** |
| Life satisfaction | 0.10 | 0.14 | 0.08 | 0.12 |
| Loneliness x (-1) | 0.10 | 0.14 | 0.04 | 0.06 |
| Depressed x (-1) | 0.10 | 0.14 | 0.04 | 0.06 |
| Anxious x (-1) | 0.10 | 0.14 | 0.23 | 0.34 |
| Absorbed | 0.10 | 0.14 | 0.18 | 0.28 |
| Bored x (-1) | 0.10 | 0.14 | 0.24 | 0.37 |
| SMS happiness | 0.10 | 0.14 | 0.10 | 0.15 |
| SMS positive emotion | 0.10 | 0.14 | 0.11 | 0.17 |
| SMS not lonely | 0.10 | 0.14 | 0.08 | 0.12 |
| **Total sum** | 1.00 | **1.41** | 1.00 | **1.52** |

Notes: The table lists the components of the "Subjective well-being summary index" defined in Allcott, Gentzkow, and Song (2022) and, for each of the four summary-index procedures studied, the implicit component weights computed as in Table II. The final row reports the total weight for each summary index. Boldface indicates negative weights and total weights greater than 1.

TABLE V

POLITICAL POLARIZATION SUMMARY INDEX: COMPONENTS AND WEIGHTS

| Component | **SN** no std | **SN** std | **IC** no std | **IC** std |
|---|---|---|---|---|
| Party affective polarization | 0.14 | 0.21 | **−0.09** | **−0.15** |
| Trump affective polarization | 0.14 | 0.21 | 0.13 | 0.20 |
| Party anger | 0.14 | 0.21 | 0.20 | 0.32 |
| Congenial news exposure | 0.14 | 0.21 | 0.27 | 0.43 |
| Issue polarization | 0.14 | 0.21 | 0.15 | 0.24 |
| Belief polarization | 0.14 | 0.21 | 0.24 | 0.38 |
| Vote polarization | 0.14 | 0.21 | 0.11 | 0.17 |
| **Total sum** | 1.00 | **1.49** | 1.00 | **1.60** |

Notes: The table lists the components of the "Political polarization summary index" defined in Allcott, Gentzkow, and Song (2022) and, for each of the four summary-index procedures studied, the implicit component weights computed as in Table II. The final row reports the total weight for each summary index. Boldface indicates negative weights and total weights greater than 1.

### 4.2. EXAMPLE B: EFFECT OF COGNITIVE BEHAVIORAL THERAPY (BARANOV ET AL. (2020A))

The article studies the effects of cognitive behavioral therapy on women's mental health, financial empowerment and decision-making, parental investment in children, and child development. It analyzes a cluster-randomized controlled trial (the Thinking Healthy Program) in rural Punjab, Pakistan, in which 903 women clinically diagnosed with prenatal depression were randomized to receive cognitive behavioral therapy. The article constructs multiple summary indices (depression severity index, financial empowerment index, parental investment summary index, etc.) using a variant of the **IC** procedure.[7] Using least-squares estimates with summary indices as left-hand-side variables, the study finds positive and persistent effects of the therapy on mental health and other areas of women's life, such as financial empowerment and parental investments.

The depression severity summary index at the 1-year follow-up was computed from a depression indicator, the depression severity (Hamilton) score, the BDQ disability score, and the Global Assessment of Functioning (GAF) score, where the latter component receives a negative weight, as reported in Table VI. For the depression severity index at the 6-month follow-up, the **IC** summary index as described in Procedure 2 does not produce a negative weight. However, due to the way the paper treats missing values when computing the summary index, their procedure does produce a negative weight for the depression severity (HAM-D) component.

The study also constructs a depression trajectory index from depression severity measured at the 6-month, 1-year, and 7-year follow-ups. As reported in Table VII, some components of the 6-month depression severity index that previously did not receive negative weights begin to do so once included in this trajectory index.

[7]Because of the way the paper treats missing values, the implicit weights their procedure assigns to components are not equal to those resulting from Procedure 2.

TABLE VI

DEPRESSION SEVERITY INDEX AT 1 YEAR FOLLOW-UP: COMPONENTS AND WEIGHTS

| Component | **SN** no std | **SN** std | **IC** no std | **IC** std |
|---|---|---|---|---|
| Depressed 1y | 0.25 | 0.26 | 0.27 | 0.28 |
| Depression severity 1y (HAM-D) | 0.25 | 0.26 | 0.21 | 0.22 |
| BDQ disability score 1y | 0.25 | 0.26 | 0.60 | 0.63 |
| General functioning 1y (GAF) | 0.25 | 0.26 | **−0.09** | **−0.09** |
| **Total sum** | 1.00 | **1.05** | 1.00 | **1.04** |

Notes: This table reports the weights assigned to each component of the 1-year depression severity index by four summary index procedures, following Baranov et al. (2020a). **SN** no std and **SN** std are scale-normalized (equal-weight) procedures without and with final index standardization, respectively. **IC** no std and **IC** std are inverse-covariance-weighted procedures without and with final index standardization.

TABLE VII

MATERNAL DEPRESSION TRAJECTORY INDEX: COMPONENTS AND WEIGHTS

| Component | **SN** no std | **SN** std | **IC** no std | **IC** std |
|---|---|---|---|---|
| Depressed (7y) | 0.08 | 0.12 | 0.12 | 0.20 |
| # Depression symptoms present (7y) | 0.08 | 0.12 | 0.01 | 0.01 |
| Symptoms cause impairment (7y) | 0.08 | 0.12 | 0.14 | 0.23 |
| Depressed in previous 2 years (7y) | 0.08 | 0.12 | 0.32 | 0.53 |
| Depressed (1y) | 0.08 | 0.12 | 0.12 | 0.20 |
| Depression severity 1y (Hamilton score) | 0.08 | 0.12 | 0.07 | 0.12 |
| BDQ disability score 1y | 0.08 | 0.12 | 0.09 | 0.16 |
| General functioning (GAF) 1y | 0.08 | 0.12 | **−0.11** | **−0.19** |
| Depressed 6m | 0.08 | 0.12 | 0.14 | 0.23 |
| Depression severity 6m (Hamilton score) | 0.08 | 0.12 | **−0.01** | **−0.02** |
| BDQ disability score 6m | 0.08 | 0.12 | 0.16 | 0.26 |
| General functioning (GAF) 6m | 0.08 | 0.12 | **−0.05** | **−0.09** |
| **Total sum** | 1.00 | **1.47** | 1.00 | **1.64** |

Notes: This table reports the weights assigned to each depression trajectory component by four summary index procedures, following Baranov et al. (2020a). **SN** no std and **SN** std are scale-normalized (equal-weight) procedures without and with final index standardization, respectively. **IC** no std and **IC** std are inverse-covariance-weighted procedures without and with final index standardization.

### 4.3. Example C: Effect of Skill Certification (Carranza et al. (2022))

The article studies whether providing additional information about workseekers' skills – and specifically, which side of the market receives it and how credibly it can be shared – affects job search behavior, firms' hiring decisions, and workseekers' employment and earnings. The study examines 6,891 young job seekers in metropolitan Johannesburg who had completed secondary school but had limited postsecondary education, work experience, and professional networks. A series of field experiments varied whether assessments of participants' skills results were disclosed privately to participants, credibly shared with employers, or provided only to firms. I refer to these treatment arms as private certification, public certification, and placebo certification, respectively, following the labels used in the replication files.

The researchers collect several employment-related outcomes, including earnings measures, labor-contract types, employment status at different points in time, and hours worked. These outcomes are used to construct several employment indices, some focusing on earnings and others on employment status and time spent working, as well as a broad labor market summary index that aggregates metrics across all three dimensions. All summary indices are constructed using the **IC** procedure. The components of the endline employment index and their weights are reported in Table VIII; the labor market index was used as an illustration in Section 3.1, see Table III.

TABLE VIII

ENDLINE EMPLOYMENT INDEX: COMPONENTS AND WEIGHTS

| Component | **SN** no std | **SN** std | **IC** no std | **IC** std |
|---|---|---|---|---|
| Earnings (weekly) | 0.33 | 0.38 | **−0.20** | **−0.23** |
| Wage (hourly) | 0.33 | 0.38 | 0.65 | 0.77 |
| Formal work | 0.33 | 0.38 | 0.54 | 0.64 |
| **Total sum** | 1.00 | **1.15** | 1.00 | **1.18** |

Notes: The table lists the components of the "Employment summary index" defined in Carranza et al. (2022) and, for each of the four summary-index procedures studied, the implicit component weights computed as in Table II. The final row reports the total weight for each summary index. Boldface indicates negative weights and total weights greater than 1.

### 4.4. Local Power Comparison

Table IX reports plug-in local powers computed from empirical treatment-effect estimates and covariance matrices. These calculations replace the population quantities in the local-power formulas for index-based tests (Proposition 6 in Section 3.3.1) and for Bonferroni-corrected component tests (Proposition C.1 in Appendix C) with sample analogues, and should be read as numerical illustrations of the formulas rather than as estimates of finite-sample power in these applications. In the Facebook-deactivation and skill-certification applications, I use the estimated treatment-effect vectors without rescaling. In the maternal-depression application, the unscaled estimates are large relative to their standard errors, yielding local powers essentially equal to one. I therefore divide the estimated component treatment-effect vector by four before computing power; this preserves the estimated direction and reduces only its magnitude.

TABLE IX

LOCAL-POWER COMPARISONS USING EMPIRICAL ESTIMATES

| Application | Index | **SN** | **IC** | Bonferroni | Ranking |
|---|---|---|---|---|---|
| Allcott et al. | Subjective well-being | 0.563 | 0.597 | 0.704 | Bonferroni > **IC** > **SN** |
| Allcott et al. | Political polarization | 0.704 | 0.929 | 0.949 | Bonferroni > **IC** > **SN** |
| Baranov et al. | Depression severity 1y | 0.435 | 0.415 | 0.334 | **SN** > **IC** > Bonferroni |
| Baranov et al. | Depression trajectory | 0.579 | 0.513 | 0.387 | **SN** > **IC** > Bonferroni |
| Carranza et al. | Employment index, public certification | 0.984 | 0.950 | 0.988 | Bonferroni > **SN** > **IC** |
| Carranza et al. | Employment index, private certification | 0.608 | 0.557 | 0.520 | **SN** > **IC** > Bonferroni |
| Carranza et al. | Labor market index, public certification | 0.989 | 0.913 | 0.990 | Bonferroni > **SN** > **IC** |
| Carranza et al. | Labor market index, placebo certification | 0.078 | 0.062 | 0.049 | **SN** > **IC** > Bonferroni |

Notes: The table reports plug-in local power at $\alpha = 0.05$. For the maternal-depression application, the empirical component treatment-effect vector is divided by four before applying the local-power formula. All other applications use the empirical estimates without rescaling.

Read as numerical illustrations of the local-power formulas, the entries of Table IX show no dominant procedure: the Bonferroni rule has the highest plug-in local power in four of the eight cases and the **SN** summary index in the other four, while the **IC** summary index never does. This mirrors the theoretical results in Section 3.3: no summary index has uniformly highest local power, and the **IC** summary index is optimal under Assumption 5 with equal standardized effects (Proposition 7).

## 5. CONCLUSION

To conclude, for the estimators and summary index constructions used in practice, the estimated effect on a summary index is a weighted sum of the estimated effects on its components, with weights that are implicit in the construction. For the **SN** summary index without final standardization the weights are equal and the estimate is the average control-group-standardized component effect; for the **IC** summary index the weights are unrestricted in sign and magnitude, and the illustrations in Section 4 show that negative weights arise in real datasets. Valid inference on the index effect is available in two forms: confidence intervals from the plug-in variance estimator in Appendix B, which accounts for the variability of the weights, and tests of hypotheses about the component effects from the shifted t-statistic of Proposition 5, which needs no such correction; the conventional t-test of the null of no effect is the special case $\tau_0 = 0$. The settings in which summary indices are commonly used do not guarantee the improvement in statistical power that is often claimed for them; Proposition 7 gives conditions under which it holds.

These results suggest the following practice.

1. For the **SN** summary index without final standardization the weights on the standardized component effects are $1/p$ and need not be reported. For constructions whose weights depend on the data, in particular **IC** summary indices, compute and report the weights (Table II). A summary index that assigns a negative weight to a component should not be used: its estimated effect cannot be interpreted as a summary of the component effects.
2. Omit the final standardization of the summary index. It rescales the average standardized component effect by a factor that depends only on the correlation structure of the components, and thereby removes comparability with component-level effects without adding information.
3. When component treatment effects are reported, construct confidence intervals for them from the joint asymptotic normality of the component estimates: by Assumption 2, $\sqrt{n}(\hat{\tau}_n - \tau) \xrightarrow{d} \mathcal{N}(0, \mathrm{Var}(\vartheta(W)))$, and the influence functions, $\vartheta(\cdot)$, are known for the estimators used in practice, so the covariance matrix can be estimated. Such intervals need be neither conservative nor based on resampling.
4. Use the conventional t-statistic of the summary index regression only to test the null of no effect on any component; for confidence intervals, or for testing a nonzero effect on the summary index, use the corrected variance in Appendix B.

5. When components are missing for some observations, do not renormalize the weights over the observed components (Appendix E); estimate the component effects under explicit missing-data assumptions and combine them with fixed weights.

## REFERENCES

Alfonsi, L., O. Bandiera, V. Bassi, R. Burgess, I. Rasul, M. Sulaiman, and A. Vitali. (2020): "Tackling youth unemployment: Evidence from a labor market experiment in Uganda," *Econometrica,* 88, 2369–2414.

Allcott, H., L. Braghieri, S. Eichmeyer, and M. Gentzkow. (2020a): "The welfare effects of social media," *American Economic Review,* 110, 629–76.

Allcott, H., L. Braghieri, S. Eichmeyer, and M. Gentzkow. (2020b): "Replication Archive for: The Welfare Effects of Social Media," American Economic Association, https://doi.org/10.3886/E112081V1.

Allcott, H., M. Gentzkow, and L. Song. (2022): "Digital addiction," *American Economic Review,* 112, 2424–63.

Anderson, M. L. (2008): "Multiple inference and gender differences in the effects of early intervention: A reevaluation of the Abecedarian, Perry Preschool, and Early Training Projects," *Journal of the American Statistical Association,* 103, 1481–95.

Anderson, T. W. (1963): "Asymptotic theory for principal component analysis," *The Annals of Mathematical Statistics,* 34, 122–48.

Asher, S., and P. Novosad. (2020): "Rural roads and local economic development," *American Economic Review,* 110, 797–823.

Banerjee, A., E. Duflo, R. Glennerster, and C. Kinnan. (2015): "The miracle of microfinance? Evidence from a randomized evaluation," *American Economic Journal: Applied Economics,* 7, 22–53.

Baranov, V., S. Bhalotra, P. Biroli, and J. Maselko. (2020a): "Maternal depression, women's empowerment, and parental investment: Evidence from a randomized controlled trial," *American Economic Review,* 110, 824–59.

Baranov, V., S. Bhalotra, P. Biroli, and J. Maselko. (2020b): "Data and Code for: Maternal Depression, Women's Empowerment, And Parental Investment: Evidence from a Randomized Control Trial," American Economic Association, https://doi.org/10.3886/E111366V1.

Baseler, T., T. Ginn, R. Hakiza, H. Ogude-Chambert, and O. Woldemikael. (2025): "Can redistribution change policy views? Aid and attitudes toward refugees," *Journal of Political Economy,* 133.

Beraja, M., A. Kao, D. Y. Yang, and N. Yuchtman. (2023): "AI-tocracy," *The Quarterly Journal of Economics,* 138, 1349–1402.

Bessone, P., G. Rao, F. Schilbach, H. Schofield, and M. Toma. (2021): "The economic consequences of increasing sleep among the urban poor," *The Quarterly Journal of Economics,* 136, 1887–1941.

Bhatt, M. P., S. B. Heller, M. Kapustin, M. Bertrand, and C. Blattman. (2024): "Predicting and preventing gun violence: An experimental evaluation of READI Chicago," *The Quarterly Journal of Economics,* 139, 1–56.

Blattman, C., J. C. Jamison, and M. Sheridan. (2017): "Reducing crime and violence: Experimental evidence from cognitive behavioral therapy in Liberia," *American Economic Review,* 107, 1165–1206.

Borusyak, K., X. Jaravel, and J. Spiess. (2024): "Revisiting event-study designs: robust and efficient estimation," *Review of Economic Studies,* 91, 3253–85.

Braghieri, L., R. Levy, and A. Makarin. (2022): "Social media and mental health," *American Economic Review,* 112, 3660–93.

Bugni, F. A., I. A. Canay, and S. McBride. (2023): "Decomposition and interpretation of treatment effects in settings with delayed outcomes," *arXiv preprint arXiv:2302.11505,*.

Bursztyn, L., A. L. González, and D. Yanagizawa-Drott. (2020): "Misperceived social norms: Women working outside the home in Saudi Arabia," *American Economic Review,* 110, 2997–3029.

Callaway, B., and P. H. Sant'Anna. (2021): "Difference-in-differences with multiple time periods," *Journal of Econometrics,* 225, 200–230.

Cantoni, D., Y. Chen, D. Y. Yang, N. Yuchtman, and Y. J. Zhang. (2017): "Curriculum and ideology," *Journal of Political Economy,* 125, 338–92.

Cantoni, D., D. Y. Yang, N. Yuchtman, and Y. J. Zhang. (2019): "Protests as strategic games: experimental evidence from Hong Kong's antiauthoritarian movement," *The Quarterly Journal of Economics,* 134, 1021–77.

Carranza, E., R. Garlick, K. Orkin, and N. Rankin. (2022): "Job search and hiring with limited information about workseekers' skills," *American Economic Review,* 112, 3547–83.

Carranza, E., R. Garlick, K. Orkin, and N. Rankin. (2023): "Data and Code for: Job Search and Hiring with Limited Information About Workseekers' Skills," American Economic Association, https://doi.org/10.3886/E172902V2.

Casey, K., R. Glennerster, and E. Miguel. (2012): "Reshaping institutions: Evidence on aid impacts using a preanalysis plan," *The Quarterly Journal of Economics,* 127, 1755–1812.

Chen, Y., and D. Y. Yang. (2019): "The impact of media censorship: 1984 or brave new world?," *American Economic Review,* 109, 2294–2332.

Chetty, R., J. N. Friedman, N. Hilger, E. Saez, D. W. Schanzenbach, and D. Yagan. (2011): "How does your kindergarten classroom affect your earnings? Evidence from Project STAR," *The Quarterly Journal of Economics,* 126, 1593–1660.

Christensen, D., O. Dube, J. Haushofer, B. Siddiqi, and M. Voors. (2021): "Building resilient health systems: experimental evidence from Sierra Leone and the 2014 Ebola outbreak," *The Quarterly Journal of Economics,* 136, 1145–98.

Currie, J., L. Davis, M. Greenstone, and R. Walker. (2015): "Environmental health risks and housing values: evidence from 1,600 toxic plant openings and closings," *American Economic Review,* 105, 678–709.

de Chaisemartin, C., and X. d'Haultfœuille. (2020): "Two-way fixed effects estimators with heterogeneous treatment effects," *American Economic Review,* 110, 2964–96.

Egger, D., J. Haushofer, E. Miguel, P. Niehaus, and M. Walker. (2022): "General equilibrium effects of cash transfers: experimental evidence from Kenya," *Econometrica,* 90, 2603–43.

Filmer, D., and L. H. Pritchett. (2001): "Estimating wealth effects without expenditure data —or tears: an application to educational enrollments in states of India," *Demography,* 38, 115–32.

Finkelstein, A., S. Taubman, B. Wright, M. Bernstein, J. Gruber, J. P. Newhouse, H. Allen, K. Baicker, and the Oregon Health Study Group. (2012): "The Oregon health insurance experiment: evidence from the first year," *The Quarterly Journal of Economics,* 127, 1057–1106.

Goldsmith-Pinkham, P., P. Hull, and M. Kolesár. (2024): "Contamination bias in linear regressions," *American Economic Review,* 114, 4015–51.

Goodman-Bacon, A. (2021): "Difference-in-differences with variation in treatment timing," *Journal of Econometrics,* 225, 254–77.

Haushofer, J., and J. Shapiro. (2016): "The short-term impact of unconditional cash transfers to the poor: experimental evidence from Kenya," *The Quarterly Journal of Economics,* 131, 1973–2042.

Hawkins, A., C. Hollrah, S. Miller, L. R. Wherry, G. Aldana, and M. Wong. (2025): "The long-term effects of income for at-risk infants: Evidence from Supplemental Security Income," *American Economic Review,* 115, 3081–3129.

Hotelling, H. (1931): "The generalization of Student's ratio," *The Annals of Mathematical Statistics,* 2, 360–78.

Hoynes, H., D. W. Schanzenbach, and D. Almond. (2016): "Long-run impacts of childhood access to the safety net," *American Economic Review,* 106, 903–34.

Jöreskog, K. G., and A. S. Goldberger. (1975): "Estimation of a model with multiple indicators and multiple causes of a single latent variable," *Journal of the American Statistical Association,* 70, 631–39.

Kinnan, C., K. Samphantharak, R. Townsend, and D. Vera-Cossio. (2024): "Propagation and insurance in village networks," *American Economic Review,* 114, 252–84.

Kling, J. R., J. B. Liebman, and L. F. Katz. (2007): "Experimental analysis of neighborhood effects," *Econometrica,* 75, 83–119.

Lau, C. P. (2026): "Aggregating Treatment Effects across Multiple Outcomes,", https://conroylau.github.io/conroy_lau_jmp.pdf.

Levy, R. (2021): "Social media, news consumption, and polarization: Evidence from a field experiment," *American Economic Review,* 111, 831–70.

List, J. A., A. M. Shaikh, and Y. Xu. (2019): "Multiple hypothesis testing in experimental economics," *Experimental Economics,* 22, 773–93.

Mogstad, M., A. Santos, and A. Torgovitsky. (2018): "Using instrumental variables for inference about policy relevant treatment parameters," *Econometrica,* 86, 1589–619.

O'Brien, P. C. (1984): "Procedures for comparing samples with multiple endpoints," *Biometrics,* 40, 1079–87.

Ramos-Toro, D. (2023): "Social exclusion and social preferences: Evidence from Colombia's leper colony," *American Economic Review,* 113, 1294–1333.

Romano, J. P., and M. Wolf. (2005): "Exact and approximate stepdown methods for multiple hypothesis testing," *Journal of the American Statistical Association,* 100, 94–108.

Schwab, B., S. Janzen, N. P. Magnan, and W. M. Thompson. (2020): "Constructing a summary index using the standardized inverse-covariance weighted average of indicators," *The Stata Journal,* 20, 952–64.

Sherman, J., and W. J. Morrison. (1950): "Adjustment of an inverse matrix corresponding to a change in one element of a given matrix," *The Annals of Mathematical Statistics,* 21, 124–27.

Stantcheva, S. (2021): "Understanding tax policy: How do people reason?," *The Quarterly Journal of Economics,* 136, 2309–69.

Sun, L., and S. Abraham. (2021): "Estimating dynamic treatment effects in event studies with heterogeneous treatment effects," *Journal of Econometrics,* 225, 175–99.

Tyler, D. E. (1981): "Asymptotic inference for eigenvectors," *The Annals of Statistics,* 9, 725–36.

van der Vaart, A. W. (2000): *Asymptotic Statistics,* Cambridge University Press.

WESTFALL, P. H., AND S. S. YOUNG. (1993): *Resampling-Based Multiple Testing: Examples and Methods for P-Value Adjustment,* John Wiley & Sons.

A. PROOFS AND AUXILIARY RESULTS

PROPOSITION A.1 — Derivation of the weights in Table II: *Let $\hat{S}_0$ be the diagonal matrix of control-group sample standard deviations, let $\hat{\Sigma}$ and $\hat{\Sigma}_0$ be the full-sample and control-group sample covariance matrices of $\boldsymbol{y}$, and define $\hat{\boldsymbol{\tau}}_{n,\mathrm{st}} := \hat{S}_0^{-1}\hat{\boldsymbol{\tau}}_n$. For the four summary index procedures in Table II, the weight vectors on $\hat{\boldsymbol{\tau}}_{n,\mathrm{st}}$ are those reported in Table II.*

PROOF *As discussed in Section 3.1, demeaning in Step 2 only changes the constant term in the affine representation of the summary index. This constant cancels from the estimator because $\frac{1}{n}\sum_{i=1}^{n}\omega(\tilde{x}_i,\hat{\nu}_n)=0$. Hence the proof can ignore demeaning and derive only the row vector $A_n$ such that $s_i = A_n\boldsymbol{y}_i + b_n$; the weight vector in Table II is then $A_n\hat{S}_0$, since $\hat{\boldsymbol{\tau}}_n = \hat{S}_0\hat{\boldsymbol{\tau}}_{n,\mathrm{st}}$.*

*For the* ***SN*** *procedure without final standardization,*

$$s_i = p^{-1}\vec{1}_p'\hat{S}_0^{-1}\boldsymbol{y}_i + b_n,\ A_n^{\mathbf{SN}} = p^{-1}\vec{1}_p'\hat{S}_0^{-1}.$$

*Therefore $A_n^{\mathbf{SN}}\hat{S}_0 = p^{-1}\vec{1}_p'$, which gives the first row of Table II.*

*If the* ***SN*** *index is standardized one more time, the control-group variance of the unstandardized index equals*

$$p^{-2}\vec{1}_p'\hat{S}_0^{-1}\hat{\Sigma}_0\hat{S}_0^{-1}\vec{1}_p.$$

*Dividing by its square root gives*

$$A_n^{\mathbf{SN},\,\mathrm{st}} = \left(\vec{1}_p'\hat{S}_0^{-1}\hat{\Sigma}_0\hat{S}_0^{-1}\vec{1}_p\right)^{-\frac{1}{2}}\vec{1}_p'\hat{S}_0^{-1},$$

*and hence*

$$A_n^{\mathbf{SN},\,\mathrm{st}}\hat{S}_0 = \left(\vec{1}_p'\hat{S}_0^{-1}\hat{\Sigma}_0\hat{S}_0^{-1}\vec{1}_p\right)^{-\frac{1}{2}}\vec{1}_p',$$

*which is the second row.*

*For the* ***IC*** *procedure, let $\tilde{\boldsymbol{y}}_i = \hat{S}_0^{-1}\boldsymbol{y}_i$ after dropping irrelevant demeaning constants. The covariance matrix of $\tilde{\boldsymbol{y}}$ is*

$$\tilde{V}_n = \hat{S}_0^{-1}\hat{\Sigma}\hat{S}_0^{-1},\ \tilde{V}_n^{-1} = \hat{S}_0\hat{\Sigma}^{-1}\hat{S}_0.$$

*Therefore, without final standardization,*

$$s_i = \left(\vec{1}_p' \tilde{V}_n^{-1} \vec{1}_p\right)^{-1} \vec{1}_p' \tilde{V}_n^{-1} \tilde{\boldsymbol{y}}_i = A_n^{\mathbf{IC}} \boldsymbol{y}_i + b_n,$$

*where*

$$A_n^{\mathbf{IC}} = \left(\vec{1}_p' \hat{S}_0 \hat{\Sigma}^{-1} \hat{S}_0 \vec{1}_p\right)^{-1} \vec{1}_p' \hat{S}_0 \hat{\Sigma}^{-1}.$$

*Multiplying by $\hat{S}_0$ gives the third row of Table II:*

$$A_n^{\mathbf{IC}} \hat{S}_0 = \left(\vec{1}_p' \hat{S}_0 \hat{\Sigma}^{-1} \hat{S}_0 \vec{1}_p\right)^{-1} \vec{1}_p' \hat{S}_0 \hat{\Sigma}^{-1} \hat{S}_0.$$

*Finally, if the **IC** index is standardized one more time, the scalar normalization from the previous display cancels from numerator and denominator. The control-group variance of the unnormalized linear combination $\vec{1}_p' \hat{S}_0 \hat{\Sigma}^{-1} \boldsymbol{y}_i$ is*

$$\vec{1}_p' \hat{S}_0 \hat{\Sigma}^{-1} \hat{\Sigma}_0 \hat{\Sigma}^{-1} \hat{S}_0 \vec{1}_p.$$

*Thus*

$$A_n^{\mathbf{IC},\,\mathrm{st}} = \left(\vec{1}_p' \hat{S}_0 \hat{\Sigma}^{-1} \hat{\Sigma}_0 \hat{\Sigma}^{-1} \hat{S}_0 \vec{1}_p\right)^{-\frac{1}{2}} \vec{1}_p' \hat{S}_0 \hat{\Sigma}^{-1},$$

*and hence*

$$A_n^{\mathbf{IC},\,\mathrm{st}} \hat{S}_0 = \left(\vec{1}_p' \hat{S}_0 \hat{\Sigma}^{-1} \hat{\Sigma}_0 \hat{\Sigma}^{-1} \hat{S}_0 \vec{1}_p\right)^{-\frac{1}{2}} \vec{1}_p' \hat{S}_0 \hat{\Sigma}^{-1} \hat{S}_0,$$

*which is the fourth row of Table II.* *Q.E.D.*

PROOF *of Proposition 3* *By items 1 and 2 of Assumption 1, $\Sigma := \mathrm{Var}[\boldsymbol{y}] = \lambda\lambda' \,\mathrm{Var}(F) + \Psi$ with $\Psi$ diagonal, so by the Sherman–Morrison formula (Sherman and Morrison (1950))*

$$\Sigma^{-1} = \Psi^{-1} - \frac{\mathrm{Var}(F)\, \Psi^{-1} \lambda\lambda' \Psi^{-1}}{1 + \mathrm{Var}(F)\, \lambda' \Psi^{-1} \lambda}.$$

*By items 2 and 3, $\mathrm{Var}_0(y_j) = \lambda_j^2 \,\mathrm{Var}_0(F) + \sigma_j^2$, so the diagonal matrix of control-group standard deviations is*

$$S_0 = \mathrm{Diag}\left(\sqrt{\mathrm{Var}_0(y_1)}, \ldots, \sqrt{\mathrm{Var}_0(y_p)}\right).$$

*Let $\underset{+}{\propto}$ denote proportionality with a positive constant. The population version of the* ***IC*** *weighting vector in Table II satisfies* $A^{\mathbf{IC}} \underset{+}{\propto} \vec{1}_p' S_0 \Sigma^{-1} S_0$. *Substituting* $\Sigma^{-1}$ *and writing* $\kappa := \left(\vec{1}_p' S_0 \Psi^{-1} \lambda\right) / (1 + \mathrm{Var}(F)\, \lambda' \Psi^{-1} \lambda)$,

$$A^{\mathbf{IC}} \underset{+}{\propto} \vec{1}_p' S_0 \Psi^{-1} S_0 - \kappa\, \mathrm{Var}(F)\, \lambda' \Psi^{-1} S_0,$$

*and the jth entry equals*

$$a_j \underset{+}{\propto} \frac{\mathrm{Var}_0(y_j)}{\sigma_j^2} - \kappa\, \mathrm{Var}(F)\, \frac{\lambda_j \sqrt{\mathrm{Var}_0(y_j)}}{\sigma_j^2} =$$

$$= \frac{\mathrm{Var}_0(y_j)}{\sigma_j^2} \left( 1 - \kappa\, \mathrm{Var}(F)\, \frac{\lambda_j}{\sqrt{\mathrm{Var}_0(y_j)}} \right).$$

*By item 4,* $\lambda \geq 0$ *and* $\lambda \neq 0$, *so* $\kappa > 0$ *since* $S_0$ *and* $\Psi^{-1}$ *are diagonal with positive entries. Hence*

$$a_j < 0 \Leftrightarrow \frac{\lambda_j}{\sqrt{\mathrm{Var}_0(y_j)}} > \frac{1}{\kappa\, \mathrm{Var}(F)} = \frac{1 + \mathrm{Var}(F)\, \lambda' \Psi^{-1} \lambda}{\mathrm{Var}(F)\, \vec{1}_p' S_0 \Psi^{-1} \lambda} =: r > 0.$$

*Q.E.D.*

PROPOSITION A.2 — Asymptotic linearity of **SN** weights: *Suppose* $\{W_i\}_{i=1}^n$ *are i.i.d.,* $D_i \in \{0,1\}$, *and* $p_0 := P(D=0) > 0$. *For every component j used in at least one* ***SN*** *summary index, let*

$$\mu_{j,0} := E\left[y_j \mid D=0\right],\ \sigma_{j,0}^2 := \mathrm{Var}\left[y_j \mid D=0\right], \tag{23}$$

*and assume* $0 < \sigma_{j,0}^2 < \infty$ *and* $E\left[\left(y_j - \mu_{j,0}\right)^4 \mid D=0\right] < \infty$.

*Let q non-standardized* ***SN*** *summary indices be constructed from nonempty sets* $\mathcal{G}_\ell \subseteq \{1,\dots,p\}$, $\ell = 1,\dots,q$, *where step 4 of Procedure 1 is omitted. Let* $A_n^{\mathrm{SN}} = \left(a_{n,\ell,j}\right)_{\ell=1,\dots,q; j=1,\dots,p}$ *be the corresponding weighting matrix, with*

$$a_{n,\ell,j} = \begin{cases} 0 & \text{if } j \notin \mathcal{G}_\ell \\ \left(|\mathcal{G}_\ell| \sqrt{\hat{\sigma}_{j,0,n}^2}\right)^{-1} & \text{if } j \in \mathcal{G}_\ell \end{cases},$$

*where $\hat{\sigma}^2_{j,0,n}$ is the control-group sample variance of component j. $A^{\mathrm{SN}}_n$ is the matrix $A_n$ of Proposition 1 for these q indices: it multiplies the unstandardized estimates $\hat{\boldsymbol{\tau}}_n$, so that $\hat{\beta}_n = A^{\mathrm{SN}}_n \hat{\boldsymbol{\tau}}_n$; the weights on the standardized estimates reported in Table II are the rows of $w_n = A^{\mathrm{SN}}_n \hat{S}_0$. Then*

$$A^{\mathrm{SN}}_n = A^{\mathrm{SN}} + \frac{1}{n}\sum_{i=1}^{n} \psi^{\mathrm{SN}}_P(W_i) + o_p\left(\frac{1}{\sqrt{n}}\right),$$

*where*

$$a_{\ell,j} = \begin{cases} 0 & \text{if } j \notin \mathcal{G}_\ell \\ \left(|\mathcal{G}_\ell|\sigma_{j,0}\right)^{-1} & \text{if } j \in \mathcal{G}_\ell \end{cases}$$

*and the components of $\psi^{\mathrm{SN}}_P(W_i)$ are*

$$\psi^{\mathrm{SN}}_P(W_i)_{\ell,j} = \begin{cases} 0 & \text{if } j \notin \mathcal{G}_\ell \\ -\frac{1}{2|\mathcal{G}_\ell|}\frac{1-D_i}{p_0}\frac{\left(y_{j,i}-\mu_{j,0}\right)^2-\sigma^2_{j,0}}{\left(\sigma^2_{j,0}\right)^{\frac{3}{2}}} & \text{if } j \in \mathcal{G}_\ell. \end{cases}$$

*In particular, $E\left[\psi^{\mathrm{SN}}_P(W)\right] = 0$ and $E\left[\left\|\psi^{\mathrm{SN}}_P(W)\right\|^2\right] < \infty$.*

PROOF *Consider an arbitrary element $(\ell, j)$ of $A^{\mathrm{SN}}_n$. If $j \notin \mathcal{G}_\ell$, then $a_{n,\ell,j} = a_{\ell,j} = 0$ and the result is immediate. Let $j \in \mathcal{G}_\ell$ and write $c_\ell := \frac{1}{|\mathcal{G}_\ell|}$.*

*Define the control-group empirical moments*

$$\hat{\theta}_{1,n} := \left(\sum_{i=1}^{n}(1-D_i)\right)^{-1} \sum_{i=1}^{n}(1-D_i)y_{j,i},$$

$$\hat{\theta}_{2,n} := \left(\sum_{i=1}^{n}(1-D_i)\right)^{-1} \sum_{i=1}^{n}(1-D_i)y^2_{j,i},$$

*and their population analogues $\theta_1 := E\left[y_j \mid D=0\right]$ and $\theta_2 := E\left[y^2_j \mid D=0\right]$. Then $\sigma^2_{j,0} = \theta_2 - \theta^2_1$ and*

$$a_{n,\ell,j} = f\left(\hat{\theta}_{1,n}, \hat{\theta}_{2,n}\right), f(x_1,x_2) := c_\ell\left(x_2 - x^2_1\right)^{-\frac{1}{2}}.$$

*By the standard expansion for conditional sample moments,*

$$\begin{pmatrix} \hat{\theta}_{1,n} - \theta_1 \\ \hat{\theta}_{2,n} - \theta_2 \end{pmatrix} = \frac{1}{n}\sum_{i=1}^{n} \frac{1-D_i}{p_0} \begin{pmatrix} y_{j,i} - \theta_1 \\ y_{j,i}^2 - \theta_2 \end{pmatrix} + o_p\left(\frac{1}{\sqrt{n}}\right).$$

*The fourth-moment and positive-variance assumptions imply that $f$ is continuously differentiable in a neighborhood of $(\theta_1, \theta_2)$ with probability approaching one. A first-order Taylor expansion gives*

$$a_{n,\ell,j} - a_{\ell,j} = (\nabla f(\theta_1, \theta_2))' \frac{1}{n}\sum_{i=1}^{n} \frac{1-D_i}{p_0} \begin{pmatrix} y_{j,i} - \theta_1 \\ y_{j,i}^2 - \theta_2 \end{pmatrix} + o_p\left(\frac{1}{\sqrt{n}}\right).$$

*Since*

$$\nabla f(\theta_1, \theta_2) = c_\ell \left(\sigma_{j,0}^2\right)^{-\frac{3}{2}} \begin{pmatrix} \theta_1 \\ -\frac{1}{2} \end{pmatrix},$$

*the summand equals*

$$-\frac{1}{2|\mathcal{G}_\ell|} \frac{1-D_i}{p_0} \frac{\left(y_{j,i} - \mu_{j,0}\right)^2 - \sigma_{j,0}^2}{\left(\sigma_{j,0}^2\right)^{\frac{3}{2}}},$$

*which is the displayed expression for $\psi_P^{\mathbf{SN}}(W_i)_{\ell,j}$. Because the number of matrix entries is fixed, applying this argument element by element proves the matrix representation. The zero mean and finite second moment of $\psi_P^{\mathbf{SN}}(W)$ follow from the displayed formula and the assumed fourth moments.* *Q.E.D.*

Proof *of Proposition 4 By Proposition 1, $\hat{\beta}_n = \hat{A}_n \hat{\boldsymbol{\tau}}_n$. The asymptotic linear representations in Assumptions 2 and 3 imply $\hat{\boldsymbol{\tau}}_n - \boldsymbol{\tau} = O_p\left(\frac{1}{\sqrt{n}}\right)$ and $\hat{A}_n - A = O_p\left(\frac{1}{\sqrt{n}}\right)$. Therefore,*

$$\sqrt{n}\left(\hat{\beta}_n - \beta\right) = \sqrt{n}\left(\hat{A}_n \hat{\boldsymbol{\tau}}_n - A\boldsymbol{\tau}\right)$$

$$= A\sqrt{n}(\hat{\boldsymbol{\tau}}_n - \boldsymbol{\tau}) + \sqrt{n}\left(\hat{A}_n - A\right)\boldsymbol{\tau} + \sqrt{n}\left(\hat{A}_n - A\right)(\hat{\boldsymbol{\tau}}_n - \boldsymbol{\tau}).$$

*The last term is $o_p(1)$ because it is $\sqrt{n} O_p\left(\frac{1}{\sqrt{n}}\right) O_p\left(\frac{1}{\sqrt{n}}\right)$. Substituting the two asymptotic linear representations gives*

$$\sqrt{n}\Big(\hat{\beta}_n - \beta\Big) = \frac{1}{\sqrt{n}} \sum_{i=1}^{n} (A\,\vartheta(W_i) + \psi(W_i)\,\boldsymbol{\tau}) + o_p(1).$$

*The summand has mean zero and finite second moment by Assumptions 2 and 3. The multivariate Lindeberg-Levy central limit theorem therefore implies*

$$\sqrt{n}\Big(\hat{\beta}_n - \beta\Big) \xrightarrow{d} \mathcal{N}(0, \mathrm{Var}(A\,\vartheta(W) + \psi(W)\,\boldsymbol{\tau})).$$

*This is (14). Consistency follows from the same expansion, since* $\hat{\beta}_n - \beta = O_p\Big(\frac{1}{\sqrt{n}}\Big)$*.* 
*Q.E.D.*

Proof *of Proposition 5 Write* $\hat{A}_n(\hat{\boldsymbol{\tau}}_n - \boldsymbol{\tau}_0) = A(\hat{\boldsymbol{\tau}}_n - \boldsymbol{\tau}_0) + \Big(\hat{A}_n - A\Big)(\hat{\boldsymbol{\tau}}_n - \boldsymbol{\tau}_0)$*. Under* $H_0$*,* $\boldsymbol{\tau}_0 = \boldsymbol{\tau}$*, so by Assumption 2* $\sqrt{n}(\hat{\boldsymbol{\tau}}_n - \boldsymbol{\tau}_0) = O_p(1)$ *and the second term times* $\sqrt{n}$ *is* $o_p(1) \cdot O_p(1) = o_p(1)$*. Hence*

$$\sqrt{n}\hat{A}_n(\hat{\boldsymbol{\tau}}_n - \boldsymbol{\tau}_0) = \frac{1}{\sqrt{n}} \sum_{i=1}^{n} A\,\vartheta(W_i) + o_p(1) \xrightarrow{d} \mathcal{N}(0, \mathrm{Var}(A\,\vartheta(W)))$$

*by the Lindeberg–Lévy central limit theorem, since* $E[\vartheta(W)] = 0$ *and* $E[\|\vartheta(W)\|^2] < \infty$*. Slutsky's theorem with* $\hat{V}_n \xrightarrow{p} \mathrm{Var}(A\,\vartheta(W)) > 0$ *gives* $T_n(\boldsymbol{\tau}_0) \xrightarrow{d} \mathcal{N}(0,1)$*. The second equality in (16) is* $\hat{\beta}_n = \hat{A}_n\hat{\boldsymbol{\tau}}_n$ *(Proposition 1), and for* $\boldsymbol{\tau}_0 = \mathbf{0}$ *the statistic is* $\sqrt{n}\hat{\beta}_n/\sqrt{\hat{V}_n}$*, the t-statistic of the summary index regression when* $\hat{V}_n/n$ *is its reported squared standard error.* *Q.E.D.*

Proposition A.3 — Variance decomposition for the **SN** summary index: *Let the conditions of Proposition A.2 hold with* $q = 1$ *and* $\mathcal{G}_1 = \{1, ..., p\}$*: in particular,* $D \in \{0,1\}$*,* $p_0 := P(D = 0) > 0$*, and* $\sigma_{j,0}^2 := \mathrm{Var}\big[y_j \mid D = 0\big]$*, so that* $A = p^{-1}\vec{1}_p' S_0^{-1}$ *with* $S_0 :=$ $\mathrm{Diag}\big(\sigma_{1,0}, ..., \sigma_{p,0}\big)$*. Let the influence function in Assumption 2 have the product form* $\vartheta(W) = \boldsymbol{u}\,\omega(\tilde{x}, \nu)$ *for a p-vector* $\boldsymbol{u}$*, as in (11). Write* $E_0[\cdot] := E[\cdot \mid D = 0]$*,* $\boldsymbol{\mu}_0 := E_0[\boldsymbol{y}]$*,* $\boldsymbol{z} := S_0^{-1}(\boldsymbol{y} - \boldsymbol{\mu}_0)$*,* $\tilde{\boldsymbol{u}} := S_0^{-1}\boldsymbol{u}$*,* $\boldsymbol{\tau}_{\mathrm{st}} := S_0^{-1}\boldsymbol{\tau}$*,* $\boldsymbol{z} \circ \boldsymbol{z} := \big(z_1^2, ..., z_p^2\big)'$*, and*

$$M := E_0\Big[\omega(\tilde{x}, \nu)\,\Big(\boldsymbol{z} \circ \boldsymbol{z} - \vec{1}_p\Big)\,\tilde{\boldsymbol{u}}'\Big], \quad K_0 := \mathrm{Var}[\boldsymbol{z} \circ \boldsymbol{z} \mid D = 0].$$

*Then the asymptotic variance in (14) satisfies*

$$\mathrm{Var}(A\,\vartheta(W) + \psi(W)\,\boldsymbol{\tau}) = \mathrm{Var}(A\,\vartheta(W)) + \Delta_1 + \Delta_2, \tag{24}$$

*where* $\mathrm{Var}(A\,\vartheta(W)) = p^{-2}E\Big[\omega(\tilde{x},\nu)^2\big(\vec{1}_p'\tilde{\boldsymbol{u}}\big)^2\Big]$ *is the probability limit of the conventional sandwich estimator, and*

$$\Delta_1 := 2\,\mathrm{Cov}(A\,\vartheta(W), \psi(W)\,\boldsymbol{\tau}) = -p^{-2}\,\boldsymbol{\tau}_{\mathrm{st}}' M\vec{1}_p,$$

$$\Delta_2 := \mathrm{Var}(\psi(W)\,\boldsymbol{\tau}) = \big(4p^2 p_0\big)^{-1}\,\boldsymbol{\tau}_{\mathrm{st}}' K_0\,\boldsymbol{\tau}_{\mathrm{st}}.$$

$\Delta_1$ *is linear and* $\Delta_2$ *quadratic in* $\boldsymbol{\tau}_{\mathrm{st}}$, *and* $\Delta_2 \geq 0$.

PROOF *Expanding the variance of the sum in (14) gives (24) with* $\Delta_1$ *and* $\Delta_2$ *as defined; all three terms are finite by the moment conditions of Assumption 2 and Proposition A.2. By the product form,* $A\,\vartheta(W) = p^{-1}\omega(\tilde{x},\nu)\,\vec{1}_p'\tilde{\boldsymbol{u}}$, *which gives the expression for* $\mathrm{Var}(A\,\vartheta(W))$; *that the sandwich estimator converges to it is shown in the continuation of the OLS example in Section 3.2. By Proposition A.2 with* $q = 1$ *and* $\mathcal{G}_1 = \{1, ..., p\}$, *the jth component of* $\psi(W)$ *equals* $-(2p)^{-1}(1-D)p_0^{-1}\frac{z_j^2-1}{\sigma_{j,0}}$, *hence*

$$\psi(W)\,\boldsymbol{\tau} = -(2p)^{-1}(1-D)p_0^{-1}\,\boldsymbol{\tau}_{\mathrm{st}}'\big(\boldsymbol{z}\circ\boldsymbol{z} - \vec{1}_p\big),$$

*which has mean zero because* $E_0\big[z_j^2\big] = 1$. *Using* $(1-D)^2 = 1-D$ *and* $E[(1-D)g(W)] = p_0 E_0[g(W)]$,

$$\Delta_2 = \big(4p^2p_0^2\big)^{-1}E\Big[(1-D)\Big(\boldsymbol{\tau}_{\mathrm{st}}'\big(\boldsymbol{z}\circ\boldsymbol{z} - \vec{1}_p\big)\Big)^2\Big] = \big(4p^2p_0\big)^{-1}\boldsymbol{\tau}_{\mathrm{st}}' K_0\,\boldsymbol{\tau}_{\mathrm{st}},$$

*and, both factors having mean zero,*

$$\Delta_1 = -p^{-2}p_0^{-1}E\Big[(1-D)\,\omega(\tilde{x},\nu)\,\big(\vec{1}_p'\tilde{\boldsymbol{u}}\big)\,\boldsymbol{\tau}_{\mathrm{st}}'\big(\boldsymbol{z}\circ\boldsymbol{z} - \vec{1}_p\big)\Big] = -p^{-2}\,\boldsymbol{\tau}_{\mathrm{st}}' M\vec{1}_p.$$

*Finally,* $\Delta_2 \geq 0$ *because* $K_0$ *is a covariance matrix.* *Q.E.D.*

## B. ESTIMATING STANDARD ERRORS

The asymptotic variance in (14) can be estimated by the sample covariance of estimated influence-function summands. Let

$$\xi(W_i) := A\,\vartheta(W_i) + \psi(W_i)\,\boldsymbol{\tau},$$

and let

$$\hat{\xi}_i := \hat{A}_n\,\hat{\vartheta}_i + \hat{\psi}_i\,\hat{\boldsymbol{\tau}}_n$$

be its feasible analogue, obtained by replacing the unknown population quantities with sample analogues.

Assumption 6: *Feasible influence-function estimation.* In addition to Assumptions 2 and 3, suppose

$$\hat{\boldsymbol{\tau}}_n \xrightarrow{p} \boldsymbol{\tau}, \hat{A}_n \xrightarrow{p} A,$$

and

$$\frac{1}{n}\sum_{i=1}^{n}\left\|\hat{\vartheta}_i - \vartheta(W_i)\right\|^2 \xrightarrow{p} 0, \frac{1}{n}\sum_{i=1}^{n}\left\|\hat{\psi}_i - \psi(W_i)\right\|^2 \xrightarrow{p} 0.$$

Proposition B.1 — Plug-in variance estimation: *Under Assumptions 2, 3, and 6, let*

$$\hat{V}_\beta := \frac{1}{n}\sum_{i=1}^{n}\left(\hat{\xi}_i - \bar{\hat{\xi}}\right)\left(\hat{\xi}_i - \bar{\hat{\xi}}\right)', \bar{\hat{\xi}} := \frac{1}{n}\sum_{i=1}^{n}\hat{\xi}_i.$$

*Then*

$$\hat{V}_\beta \xrightarrow{p} \mathrm{Var}(A\,\vartheta(W) + \psi(W)\,\boldsymbol{\tau}).$$

*Hence, for a scalar summary index estimate, $\sqrt{\frac{\hat{V}_\beta}{n}}$ is an asymptotically valid standard-error estimator. For $q$ summary indices, $\hat{A}_n$ is $q \times p$, $\hat{V}_\beta$ is $q \times q$, and $\sqrt{n}\left(\hat{\boldsymbol{\beta}}_n - \boldsymbol{\beta}\right) \xrightarrow{d}$ $\mathcal{N}(0, \mathrm{Var}(A\,\vartheta(W) + \psi(W)\,\boldsymbol{\tau}))$ by the argument of Proposition 4, so simultaneous confidence intervals follow from the $(1-\alpha)$-quantile of $\max_\ell |Z_\ell|$, $Z \sim \mathcal{N}\left(0, \hat{R}_n\right)$, with $\hat{R}_n$ the correlation matrix implied by $\hat{V}_\beta$.*

Proof *Let $\xi_i := \xi(W_i)$. By 6 and the consistency of $\hat{\boldsymbol{\tau}}_n$ and $\hat{A}_n$,*

$$\frac{1}{n}\sum_{i=1}^{n}\left\|\hat{\xi}_i - \xi_i\right\|^2 \xrightarrow{p} 0.$$

*Therefore the sample second moment and sample mean of $\hat{\xi}_i$ have the same probability limits as those of $\xi_i$. The weak law of large numbers gives*

$$\frac{1}{n}\sum_{i=1}^{n}\xi_i\xi_i' \xrightarrow{p} E[\xi(W)\xi'(W)], \frac{1}{n}\sum_{i=1}^{n}\xi_i \xrightarrow{p} E[\xi(W)] = 0.$$

*The displayed convergence for $\hat{V}_\beta$ follows by the continuous mapping theorem.* *Q.E.D.*

PROOF *of Proposition 6* *Let* $\boldsymbol{\tau}_n = \frac{\boldsymbol{\delta}}{\sqrt{n}}$ *and* $\beta_n = A\boldsymbol{\tau}_n$. *By Proposition 1,* $\hat{\beta}_n = \hat{A}_n\hat{\boldsymbol{\tau}}_n$. *Under Assumption 3,* $\hat{A}_n - A = O_p\left(\frac{1}{\sqrt{n}}\right)$*; under Assumption 4,* $\hat{\boldsymbol{\tau}}_n - \boldsymbol{\tau}_n = O_p\left(\frac{1}{\sqrt{n}}\right)$. *Hence*

$$\sqrt{n}\left(\hat{\beta}_n - \beta_n\right) = A\sqrt{n}(\hat{\boldsymbol{\tau}}_n - \boldsymbol{\tau}_n) + \sqrt{n}\left(\hat{A}_n - A\right)\boldsymbol{\tau}_n + \sqrt{n}\left(\hat{A}_n - A\right)(\hat{\boldsymbol{\tau}}_n - \boldsymbol{\tau}_n).$$

*Because* $\boldsymbol{\tau}_n = \frac{\boldsymbol{\delta}}{\sqrt{n}}$, *the last two terms are* $o_p(1)$. *Thus*

$$\sqrt{n}\left(\hat{\beta}_n - \beta_n\right) = \frac{1}{\sqrt{n}}\sum_{i=1}^{n} A\,\vartheta(W_i) + o_p(1) \xrightarrow{d} \mathcal{N}(0, AVA').$$

*Since* $\sqrt{n}\beta_n = A\boldsymbol{\delta}$,

$$T_n^{\mathrm{AV}} = \frac{\sqrt{n}\hat{\beta}_n}{\sqrt{AVA'}} \xrightarrow{d} \mathcal{N}(\Delta, 1),\ \Delta = \frac{A\boldsymbol{\delta}}{\sqrt{AVA'}}.$$

*Therefore, for* $z = z_{1-\alpha/2}$,

$$\pi(\Delta) = P(|Z + \Delta| > z) = \Phi(\Delta - z) + \Phi(-\Delta - z),$$

*where* $Z \sim \mathcal{N}(0, 1)$. *This proves (20).*

*It remains to characterize the maximizing direction. The function* $\pi$ *is even, and for* $x \geq 0$ *its derivative is* $f(x - z) - f(x + z) \geq 0$, *where* $f$ *is the standard normal density. Hence local power is ordered by* $|\Delta|$. *Fix* $\boldsymbol{\delta}'V^{-1}\boldsymbol{\delta} = c > 0$. *By the Cauchy–Schwarz inequality,*

$$|A\boldsymbol{\delta}|^2 = \left|\left(V^{1/2}A'\right)'\left(V^{-1/2}\boldsymbol{\delta}\right)\right|^2 \leq (AVA')(\boldsymbol{\delta}'V^{-1}\boldsymbol{\delta}).$$

*so* $\Delta^2 \leq c$, *with equality if and only if* $V^{-1/2}\boldsymbol{\delta} \propto V^{1/2}A'$, *equivalently* $\boldsymbol{\delta} \propto VA'$. *Hence, among* $\boldsymbol{\delta}$ *with* $\boldsymbol{\delta}'V^{-1}\boldsymbol{\delta} = c$, $|\Delta|$ *and therefore the local power are maximized at* $\boldsymbol{\delta} \propto VA'$. *Q.E.D.*

PROPOSITION B.2 — Assumption 6 for least squares and the **SN** summary index: *Let* $\{W_i\}_{i=1}^n$ *be i.i.d. and let* $\hat{\boldsymbol{\tau}}_n$ *be the least-squares estimator of Example 1 with* $Q := E[\tilde{x}\tilde{x}']$ *nonsingular,* $E\|\tilde{x}\|^4 < \infty$ *and* $E[\|\mathbf{u}\|^2\|\tilde{x}\|^2] < \infty$. *Let the summary index be the* ***SN*** *index without final standardization built from all p components, and let the conditions of Proposition A.2 hold. Define the feasible summands by replacing* $\left(\Gamma, Q, \mu_{j,0}, \sigma_{j,0}^2, p_0\right)$ *in* $\vartheta(W_i)$ *and* $\psi(W_i)$ *with their sample analogues* $\left(\hat{\Gamma}_n, \hat{Q}_n, \hat{\mu}_{j,0,n}, \hat{\sigma}_{j,0,n}^2, \hat{p}_{0,n}\right)$. *Then Assumption 6 holds.*

Proof *Consistency of* $\hat{\boldsymbol{\tau}}_n$ *and* $\hat{A}_n$ *follows from the weak law of large numbers and the continuous mapping theorem, since* $\hat{\Gamma}_n = \left(n^{-1}\sum_i \boldsymbol{y}_i \tilde{x}_i'\right)\left(n^{-1}\sum_i \tilde{x}_i\tilde{x}_i'\right)^{-1} \xrightarrow{p} \Gamma$ *and* $\hat{\sigma}^2_{j,0,n} \xrightarrow{p} \sigma^2_{j,0} > 0$.

*For* $\vartheta$, *write* $\hat{\vartheta}_i - \vartheta(W_i) = (\hat{\boldsymbol{u}}_i - \boldsymbol{u}_i)\tilde{x}_i'\hat{Q}_n^{-1}e_{1,k} + \boldsymbol{u}_i\tilde{x}_i'\left(\hat{Q}_n^{-1} - Q^{-1}\right)e_{1,k}$ *with* $\hat{\boldsymbol{u}}_i - \boldsymbol{u}_i = -\left(\hat{\Gamma}_n - \Gamma\right)\tilde{x}_i$. *Hence*

$$\frac{1}{n}\sum_{i=1}^{n}\left\|\hat{\vartheta}_i - \vartheta(W_i)\right\|^2 \le 2\left\|\hat{\Gamma}_n - \Gamma\right\|^2\left\|\hat{Q}_n^{-1}e_{1,k}\right\|^2\frac{1}{n}\sum_{i=1}^{n}\|\tilde{x}_i\|^4 +$$

$$+2\left\|\hat{Q}_n^{-1} - Q^{-1}\right\|^2\frac{1}{n}\sum_{i=1}^{n}\|\boldsymbol{u}_i\|^2\|\tilde{x}_i\|^2,$$

*and both terms are* $o_p(1)\cdot O_p(1)$ *by consistency of* $\hat{\Gamma}_n$ *and* $\hat{Q}_n$ *and the weak law of large numbers applied under* $E\|\tilde{x}\|^4 < \infty$ *and* $E\left[\|\boldsymbol{u}\|^2\|\tilde{x}\|^2\right] < \infty$.

*For* $\psi$, *by Proposition A.2 the jth component is* $\psi(W_i)_j = (1-D_i)h_j\left(y_{j,i};\theta\right)$ *with* $\theta := \left(p_0, \mu_{j,0}, \sigma^2_{j,0}\right)$ *and* $h_j(y;\theta) := -(2p)^{-1}p_0^{-1}\left(\left(y-\mu_{j,0}\right)^2 - \sigma^2_{j,0}\right)\left(\sigma^2_{j,0}\right)^{-\frac{3}{2}}$, *a quadratic polynomial in y whose coefficients are continuous in* $\theta$ *on* $\left\{p_0 > 0, \sigma^2_{j,0} > 0\right\}$. *Writing* $h_j(y;\theta) = a(\theta)y^2 + b(\theta)y + c(\theta)$ *and* $\hat{\theta}_n$ *for the sample analogue,*

$$\frac{1}{n}\sum_{i=1}^{n}\left(\hat{\psi}_{i,j} - \psi(W_i)_j\right)^2 \le 3\Bigg[\left(a\left(\hat{\theta}_n\right) - a(\theta)\right)^2\frac{1}{n}\sum_{i=1}^{n}(1-D_i)y_{j,i}^4 +$$

$$+\left(b\left(\hat{\theta}_n\right) - b(\theta)\right)^2\frac{1}{n}\sum_{i=1}^{n}(1-D_i)y_{j,i}^2 + \left(c\left(\hat{\theta}_n\right) - c(\theta)\right)^2\Bigg],$$

*which is* $o_p(1)$ *because* $\hat{\theta}_n \xrightarrow{p} \theta$, *the coefficients are continuous, and* $E\left[(1-D)y_j^4\right] = p_0E\left[y_j^4 \mid D=0\right] < \infty$ *by the fourth-moment condition of Proposition A.2. Summing over j gives the second condition in Assumption 6.* *Q.E.D.*

Proof *of Corollary 3.1* *Let* $w_n = A_n\hat{S}_0$ *be the row vector reported in Table II; this is the row vector multiplying the standardized estimator* $\hat{\boldsymbol{\tau}}_{n,\mathrm{st}}$. *Since* $\hat{\boldsymbol{\tau}}_n = \hat{S}_0\hat{\boldsymbol{\tau}}_{n,\mathrm{st}}$, *the row vector multiplying the unstandardized estimator* $\hat{\boldsymbol{\tau}}_n$ *is* $A_n = w_n\hat{S}_0^{-1}$. *The assumptions* $\hat{\Sigma} \xrightarrow{p} \Sigma$ *and* $\hat{\Sigma}_0 \xrightarrow{p} \Sigma_0$, *together with positive definiteness, imply by the continuous mapping theorem that* $\hat{S}_0 \xrightarrow{p} S_0$, $\hat{S}_0^{-1} \xrightarrow{p} S_0^{-1}$, *and* $\hat{\Sigma}^{-1} \xrightarrow{p} \Sigma^{-1}$.

*For **SN**, Table II gives $w_n = c_n \vec{1}_p'$ for a positive scalar $c_n$; this scalar is $p^{-1}$ without final standardization and another positive normalizing constant with final standardization. Thus*

$$A^{\mathrm{SN}} \propto \vec{1}_p' S_0^{-1}, \; \left(A^{\mathrm{SN}}\right)' \propto S_0^{-1} \vec{1}_p.$$

*Applying Proposition 6 gives*

$$\boldsymbol{\delta}_{\mathrm{SN}}^* \propto V \left(A^{\mathrm{SN}}\right)' \propto V \, S_0^{-1} \, \vec{1}_p.$$

*For **IC**, Table II gives*

$$w_n = d_n \vec{1}_p' \hat{S}_0 \hat{\Sigma}^{-1} \hat{S}_0$$

*for a positive scalar $d_n$; again, the scalar differs depending on whether the index is finally standardized. Hence*

$$A_n = w_n \hat{S}_0^{-1} = d_n \vec{1}_p' \hat{S}_0 \hat{\Sigma}^{-1},$$

*and therefore*

$$A^{\mathrm{IC}} \propto \vec{1}_p' S_0 \Sigma^{-1}, \; \left(A^{\mathrm{IC}}\right)' \propto \Sigma^{-1} S_0 \vec{1}_p,$$

*using symmetry of $S_0$ and $\Sigma$. Applying Proposition 6 gives*

$$\boldsymbol{\delta}_{\mathrm{IC}}^* \propto V \left(A^{\mathrm{IC}}\right)' \propto V \, \Sigma^{-1} \, S_0 \, \vec{1}_p.$$

*This proves the two expressions in Corollary 3.1.* *Q.E.D.*

## C. BONFERRONI-CORRECTED COMPONENT TESTS

REMARK 1: *Multiple Testing*. As an alternative to constructing summary indices, one may estimate treatment effects on the components of the summary index and test the component null hypotheses individually, correcting for multiple testing. Let $T_{j,n} := \frac{\hat{\tau}_{j,n}}{\mathrm{s.e.}(\hat{\tau}_{j,n})}$ be the component-level $t$-statistic and let $c_p := z_{1-\alpha/(2p)}$. The two-sided Bonferroni rule rejects $H_0 : \boldsymbol{\tau} = \mathbf{0}$ whenever

$$\max_{1 \le j \le p} \left| T_{j,n} \right| > c_p.$$

The following proposition gives the asymptotic local power of this rule under Assumption 4.

PROPOSITION C.1 — Local power of Bonferroni-corrected component tests: *Suppose* $\sqrt{n}\,\text{s.e.}(\hat{\tau}_{j,n})$ *is consistent for* $\sqrt{V_{j,j}}$ *for each* $j$*. Let* $\boldsymbol{T}_n := (T_{1,n}, ..., T_{p,n})'$ *and define* $V_D := \text{Diag}(V_{1,1}, ..., V_{p,p})$*. Under Assumption 4,*

$$\boldsymbol{T}_n \xrightarrow{d} \mathcal{N}\left(V_D^{-1/2}\boldsymbol{\delta}, V_D^{-1/2} V V_D^{-1/2}\right). \tag{25}$$

*Therefore, for the two-sided Bonferroni rule that rejects* $H_0 : \boldsymbol{\tau} = \mathbf{0}$ *whenever* $\max_{1\le j\le p}|T_{j,n}| > c_p$*, with* $c_p := z_{1-\alpha/(2p)}$*, the asymptotic local power equals*

$$\pi_{B(\boldsymbol{\delta},V)} := P\left(\max_{1\le j\le p}|Z_j| > c_p\right),$$

*where* $\boldsymbol{Z}$ *has the distribution in (25).*

PROOF *of Proposition C.1* *Under Assumption 4,*

$$\sqrt{n}(\hat{\boldsymbol{\tau}}_n - \boldsymbol{\tau}_n) = \frac{1}{\sqrt{n}}\sum_{i=1}^{n}\vartheta(W_i) + o_{p(1)},$$

*and* $\boldsymbol{\tau}_n = \frac{\boldsymbol{\delta}}{\sqrt{n}}$*. Hence the central limit theorem gives*

$$\sqrt{n}\hat{\boldsymbol{\tau}}_n = \boldsymbol{\delta} + \frac{1}{\sqrt{n}}\sum_{i=1}^{n}\vartheta(W_i) + o_p(1) \xrightarrow{d} \mathcal{N}(\boldsymbol{\delta}, V).$$

*Let* $V_D := \text{Diag}(V_{1,1}, ..., V_{p,p})$*. By the standard-error consistency assumed in Proposition C.1,*

$$\text{Diag}\left(\sqrt{n}\,\text{s.e.}(\hat{\tau}_{1,n}), ..., \sqrt{n}\,\text{s.e.}(\hat{\tau}_{p,n})\right) \xrightarrow{p} V_D^{\frac{1}{2}}.$$

*Slutsky's theorem therefore implies*

$$\boldsymbol{T}_n = \text{Diag}\left(\sqrt{n}\,\text{s.e.}(\hat{\tau}_{1,n}), ..., \sqrt{n}\,\text{s.e.}(\hat{\tau}_{p,n})\right)^{-1}\sqrt{n}\hat{\boldsymbol{\tau}}_n \xrightarrow{d} \mathcal{N}\left(V_D^{-\frac{1}{2}}\boldsymbol{\delta}, V_D^{-\frac{1}{2}} V V_D^{-\frac{1}{2}}\right),$$

*which is (25).*

*Let* $B := \{z \in \mathbb{R}^p : \max_{1\le j\le p}|z_j| > c_p\}$*. The boundary of* $B$ *is contained in the finite union of hyperplanes* $\{z : z_j = c_p\}$ *and* $\{z : z_j = -c_p\}$*,* $j = 1, ..., p$*. The limiting normal distribution is nondegenerate, so this boundary has probability zero. The portmanteau theorem then gives*

$$\lim_{n\to\infty} P_n\left(\max_{1\le j\le p}|T_{j,n}| > c_p\right) = P\Big(\max_{1\le j\le p}|Z_j| > c_p\Big),$$

*where $\boldsymbol{Z}$ has the distribution in (25). This is the displayed expression for $\pi_{B(\boldsymbol{\delta},V)}$. Q.E.D.*

## D. OTHER SUMMARY INDICES

The results in the main text do not depend on the literal steps in Procedure 1 or 2. They use the fact that, after the data-dependent weights are fixed, the final scalar index can be written as

$$s_{i,n} = A_n \boldsymbol{y}_i + b_n.$$

For any index with this representation, the same argument as in Proposition 1 gives $\hat{\beta}_n = A_n \hat{\boldsymbol{\tau}}_n$. The large-sample results then apply whenever $A_n$ satisfies the expansion in Assumption 3.

This covers the following common variants.

(i) *Multiple and overlapping domains*. Researchers may partition outcomes into several domains, or allow domains to overlap, and construct one index per domain, as in Casey, Glennerster, and Miguel (2012). The scalar results apply to each domain separately. Equivalently, if the $q$ indices are treated jointly, $A_n$ becomes a $q \times p$ matrix whose rows give the weights of each index on the original outcomes.

(ii) *Nested indices*. If an index is constructed from sub-indices, as in Haushofer and Shapiro (2016) and Christensen et al. (2021), the final index is still affine in the original outcomes whenever each construction step is affine. The relevant weights are the final weights on the original outcomes, not only the weights assigned to the intermediate sub-indices.

(iii) *PCA indices*. If the first principal component is used as the index, as in Filmer and Pritchett (2001), the index is linear in the standardized outcomes conditional on the estimated loading vector. The results therefore apply under regularity conditions that make this loading vector asymptotically linear (see Anderson (1963) for the Gaussian case and Tyler (1981) more generally).

(iv) *Economically weighted indices*. If outcomes are weighted by external costs or benefits, as in Bhatt et al. (2024), the representation above holds. If the weights are fixed external constants, the generated-weight term is zero; if the weights are estimated, the same results apply after accounting for their influence function.

This discussion assumes that, for each constructed index, the same row vector $A_n$ applies to every observation. Missing-value imputations that create observation-specific weights are treated separately in Appendix E.

## E. MISSING DATA

When a component of $\boldsymbol{y}$ is missing for some observation, a common practice is to construct the summary index only from the components observed for that observation, rather than requiring complete data or dropping the observation entirely. Under such an **available-components rule**, each observed component absorbs the weight that would otherwise have gone to the missing components for that observation, so a component's realized weight depends on which other components happen to be missing alongside it. Anderson (2008) motivates the missing-value adjustment used in the **IC** procedure on this basis, stating that it "uses all of the available data, but it weights outcomes with fewer missing values more heavily." This appendix formalizes the rule and shows that this intuition does not generally hold: whether a component ends up with more or less weight on average is not simply governed by how often it is missing.

The reason this requires separate treatment is that the decomposition in Proposition 1 requires a single row vector $A_n$ such that $s_{i,n} = A_n \boldsymbol{y}_i + b_n$ for every observation, which is what lets the summary-index-level effect be read off as the fixed linear combination $A_n \hat{\boldsymbol{\tau}}_n$ of component-level effects. An available-components rule generally violates this, because the weight vector it induces varies with each observation's missingness pattern.

Formally, let $m_{i,j} = 1$ if component $j$ is observed for observation $i$, and let $c_{j,n}$ denote the weight component $j$ would receive in the absence of missing data. The available-components rule renormalizes these base weights over only the components observed for $i$:

$$s_{i,n}^{\text{miss}} = \frac{\sum_{j=1}^{p} m_{i,j} c_{j,n} \boldsymbol{y}_{i,j}}{\sum_{k=1}^{p} m_{i,k} c_{k,n}}.$$

Equivalently, $s_{i,n}^{\text{miss}} = A_{i,n}^{\text{miss}} \boldsymbol{y}_i$, with observation-specific weights

$$A_{i,n,j}^{\text{miss}} = \frac{m_{i,j} c_{j,n}}{\sum_{k=1}^{p} m_{i,k} c_{k,n}}.$$

that depend on $i$ only through its missingness pattern $(m_{i,1}, \ldots, m_{i,p})$, and so generally differ across observations. Averaging these observation-specific weights over the sample gives

$$\hat{\beta}_n^{\text{miss}} = \frac{1}{n} \sum_{i=1}^{n} A_{i,n}^{\text{miss}} \boldsymbol{y}_i \omega(\tilde{x}_i, \hat{\nu}_n),$$

which is not generally equal to $A_n \hat{\boldsymbol{\tau}}_n$ for any single row vector $A_n$ applied to the component estimates in Proposition 1: no fixed set of index weights rationalizes $\hat{\beta}_n^{\text{miss}}$ as a linear combination of the component treatment effects. This is an algebraic issue separate from the missing-data assumptions needed for point identification – it arises even when missingness is unrelated to potential outcomes.

The following table quantifies why the "more weight to less-missing outcomes" intuition does not generally hold, reporting the weights induced by the equal-weight version of the available-components rule, $c_{j,n} = 1$, for three outcomes.

TABLE X

EFFECTIVE WEIGHTS UNDER AN AVAILABLE-COMPONENTS RULE

| Obs. | Observed components | $j = 1$ | $j = 2$ | $j = 3$ |
|---|---|---|---|---|
| 1 | 2 | 0 | 1 | 0 |
| 2 | 1, 3 | $\frac{1}{2}$ | 0 | $\frac{1}{2}$ |
| 3 | 1, 2 | $\frac{1}{2}$ | $\frac{1}{2}$ | 0 |
| 4 | 1, 3 | $\frac{1}{2}$ | 0 | $\frac{1}{2}$ |
| 5 | 2 | 0 | 1 | 0 |
| 6 | 1, 3 | $\frac{1}{2}$ | 0 | $\frac{1}{2}$ |
| 7 | 1, 2 | $\frac{1}{2}$ | $\frac{1}{2}$ | 0 |
| 8 | 1, 2, 3 | $\frac{1}{3}$ | $\frac{1}{3}$ | $\frac{1}{3}$ |
| Missing observations | | 2 | 3 | 4 |
| Average effective weight | | $\frac{17}{48}$ | $\frac{20}{48}$ | $\frac{11}{48}$ |

Notes: Columns $j = 1, 2, 3$ report the weights $A^{\text{miss}}_{i,n,j}$ in $s^{\text{miss}}_{i,n} = \sum_{j=1}^{3} A^{\text{miss}}_{i,n,j} \boldsymbol{y}_{i,j}$ when all observed components receive equal base weight. The final two rows report the number of missing observations and the average effective weight for each outcome.

Outcome 1 has the fewest missing observations, but its average effective weight is smaller than the average effective weight assigned to outcome 2. The reason is that the weight assigned to an observed component depends on which other components are observed for the same individual. In the **IC** procedure, covariance weighting adds another source of variation, so the effective weights are even less tightly tied to marginal missingness rates. A cleaner approach is to estimate the component effects and the index weights under explicit assumptions on the missing-data mechanism, and then form $A_n \hat{\boldsymbol{\tau}}_n$ directly.